\pdfoutput=1
\documentclass[aps,prd,11pt,superscriptaddress,notitlepage,nofootinbib]{revtex4-1}
\usepackage{microtype}
\usepackage{graphicx}
\graphicspath{{./fig/}}
\usepackage{subfigure}
\usepackage{mathtools}
\usepackage{amssymb}
\usepackage{mathrsfs}
\DeclareMathOperator{\sgn}{sgn}
\DeclarePairedDelimiter{\abs}{\lvert}{\rvert}

\usepackage[T1]{fontenc}
\usepackage[pdfusetitle,bookmarksopen,colorlinks,allcolors=blue]{hyperref}

\begin{document}
\title{Compact kink and its interaction with compact oscillons}

\author{F.~M.~Hahne}
\email{fernandomhahne@gmail.com}
\affiliation{Departamento de F\'isica, Universidade Federal de Santa Catarina, Campus Trindade, 88040-900, Florian\'opolis-SC, Brazil}

\author{P.~Klimas}
\email{pawel.klimas@ufsc.br}
\affiliation{Departamento de F\'isica, Universidade Federal de Santa Catarina, Campus Trindade, 88040-900, Florian\'opolis-SC, Brazil}

\begin{abstract}
    We study compact kinks and its interaction with compact oscillons in models with non-analytic potentials.
    Oscillon-like excitations are the main ingredients of the radiation field.
    We look at the problem of scattering which involves topological compact kinks and non-topological oscillons.
    We also look at the problem of propagation of small perturbation at the kink bulk.
\end{abstract}

\maketitle

\section{Introduction}

Scalar field models have many physical applications, ranging from fundamental  interactions through  condensed matter physics to large scale cosmological models. The localized energy configurations such as solitons or lumps that appear in scalar field theories have usually infinite sizes, which means that they cover all the  space \cite{Manton:2004tk, shnir_2018}. There are, however, some notable exceptions to this rule, i.e.\ there are models such that  their solutions are localized on a strictly finite size support. Such field  configurations are called \emph{compactons}  \cite{1993PhRvL..70..564R, PhysRevE.57.2320}. In this paper we study interaction between topological  compact kinks and non-topological compact oscillons. Compact kinks were first discussed in the field-theoretic limit of a modified system of interacting pendulums \cite{Arodz:2002yt,Arodz:2003mx}.
These kinks reach vacuum values in a parabolic way. Such approach to the vacuum is a very characteristic property of field theories with standard kinetic terms and non-standard behavior of the potential at its minima, namely, for the class of sharp potentials.
The sharpness of the potential means that its first one-sided derivatives at the minima do not vanish. The models with V-shaped potentials and standard  kinetic terms in the Lagrangian are non-unique field-theoretic models supporting compactons. Another possibility is that the potential is smooth at its minima, but the kinetic term of the model contains powers higher than two of field partial derivatives. In the simplest case, taking a square of a V-shaped potential and a square of  the kinetic term, one gets a new model that possesses a static compact kink as its Bogomolny solution. This kink has exactly the same form as the kink in the original model i.e.\ with non-analytic potential. For instance, static compact kinks presented in the model with V-shaped potential (in the case of small amplitudes) \cite{Arodz:2002yt}, and also in the model with $\phi^4$ potential \cite{Adam:2007ij} are exactly the same and given by the segment of $\sim\sin(x)$ function on $-\pi/2\le x\le\pi/2$. Of course, both these models are different, which in particular means that they have different time dependent solutions.

Models with V-shaped potentials were discussed in several physical contexts. In the systems few degrees of freedom they were investigated because they support chaotic behavior of mechanical systems \cite{PhysRevA.27.1741, PhysRevE.49.1073,PhysRevE.50.4427}. They were also discussed in plasma physics \cite{PhysRevLett.78.4761}. We would like to stress that the mechanical systems with infinite number of degrees of freedom and some rigid restrictions on values of the field have the closest relation with the models discussed in this paper. The models with non-analytic potentials can also be derived as  effective models which emerge from some other physical models through a symmetry reduction ansatz. This is exactly the case of the Skyrme model discussed in \cite{Adam:2017srx}. Its effective field equation has exactly the same form  as the Euler-Lagrange equation discussed in this paper. The analysis of small amplitude excitations presented in \cite{Adam:2017srx} leads to the conclusion that the model possesses approximated compact oscillons, i.e.\ periodic time dependent solutions that exist on a finite size support. Such field configurations, first presented in  \cite{Arodz:2007jh}, constitute a family of exact solutions of the signum-Gordon model. Some further generalizations of these solutions were discussed in \cite{Arodz:2011zm,?wierczy?ski2021}. The dynamics of compact oscillons is  discussed in  \cite{Klimas:2018woi}. In \cite{Hahne:2019ela} we pay much attention to the  scattering process which involves two compact oscillons. This process is of great importance, especially as a mechanism of production of smaller scale oscillon-like excitations. The properties of the radiation  in the signum-Gordon model and some other its modifications is a fascinating and still not fully understood subject. For instance, we have pointed out in \cite{Hahne:2019ela} that this radiation has likely a fractal nature.

In this paper we discuss the interaction of this radiation with topological compact kinks. We choose a compact oscillon as a typical ingredient of the radiation. Of course, some other exact solutions of the signum-Gordon model can also play a role in the spectrum of radiation. For instance, the model possesses infinite wave solutions \cite{Arodz:2003mx},  self-similar solutions \cite{Arodz:2007ek, Arodz:2007dp} and shock wave solutions \cite{Arodz:2005gz, Klimas:2006fs} that can partially emerge in the process of scattering of two compact structures. However,  since such solutions have infinite energy or require constant transfer of energy to sustain their form (the case of shock waves) they are less likely to emerge in the evolution of finite energy field configurations.  It is a well known fact that in non-integrable field theories (such as our model) the presence of radiation is unavoidably generated in the kink and antikink scattering process. In addition, if the model supports compactons then this radiation propagates  also on the kink support. The equation that governs this propagation strongly depends on the region in which the propagation takes place. The problem of propagation of small field excitations along the kink is a particularly interesting problem in the context of brane world scenario, like the presented in \cite{Adam:2007ij, Adam:2007ag, Adam:2008ck}, where the kink extends in an extra spatial dimension.

The paper is organized as follows.
In sections~\ref{sec:kinks} we review the solutions for static compact kinks and discuss some of its general properties.
In section~\ref{sec:shrunken-and-streched} we study the dynamics of perturbed kinks and discuss the role of oscillons as a way of emitting excess energy.
Numerical results for the process of kink-oscillon scattering are discussed in section~\ref{sec:scattering}, with the analysis of the efficiency that oscillons transverse the kink bulk as our main approach.
We expand on this topic in section~\ref{sec:perturbations} where we analyze numerically and analytically the behavior of localized perturbations inside the kink support.
Further analytical calculations for a toy-model similar to our main model are provided in section~\ref{sec:toy}.
Some remarks about a family of models that includes both our main model and toy-model are presented in section~\ref{sec:generalized}.
We review our conclusions in section~\ref{sec:conclusions}.

\section{Static compact kinks}
\label{sec:kinks}

\begin{figure}
    {\includegraphics[width=0.45\textwidth]{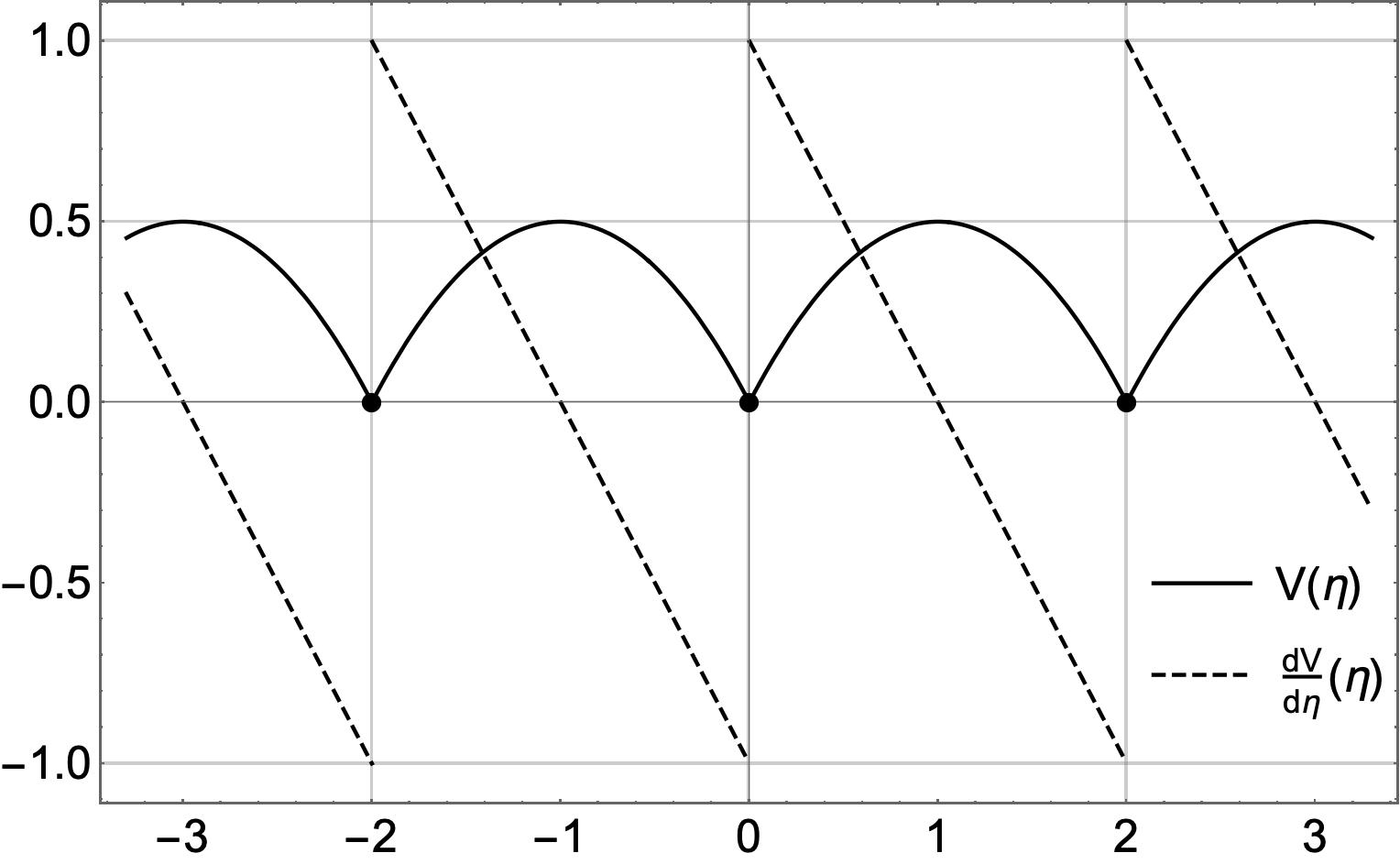}}
    \caption{The periodic potential $V(\eta)$ and its derivative $V'(\eta)$. The potential has minima at $\eta=2k$, $k=0,\pm1,\pm2,\ldots$.}
    \label{fig:VdV}
\end{figure}

We consider a scalar field theory in $1+1$ dimensions defined by the action
\begin{equation}
    {S=\int dt \, dx \left[\frac{1}{2}(\partial_t\eta)^2-\frac{1}{2}(\partial_x\eta)^2-V(\eta)\right]\label{eq0}}
\end{equation}
where $\eta(t,x)$ is a real valued scalar field. It does not matter for the purpose of this paper whether the action \eqref{eq0} is effective or not. The potential $V(\eta)$ is a periodic function which is sharp at its minima. It has the form of parabolic shapes repeated infinitely many times, see Fig.~\ref{fig:VdV}. One possible way of expressing the potential is the following
\begin{align*}
    V(\eta)   &= \sum_{n=-\infty}^{\infty} \left( |\eta - 4n| - \frac{1}{2}(\eta - 4n)^2 \right) H_n(\eta), \\
    H_n(\eta) &:= \theta(\eta - 4n + 2) - \theta(\eta - 4n - 2)
\end{align*}
where $\theta(\eta)$ is the Heaviside step function. The function $H_n(\eta)=1$ for $|\eta-4n|<2$ and $H_n(\eta)=0$ otherwise.  The potential has sharp minima at $\eta_\text{min}=\{4n,4n\pm 2\}$, $n=0\pm1,\pm2,\ldots$ or equivalently $\eta_\text{min}=2k$ where $k=0,\pm1,\pm2,\ldots$. The index $n$ labels segments of the periodic potential whereas the index $k$ labels its minima.

Models like this one can be obtained through a small-angle approximation of a system of coupled pendulums, with motion limited by rigid barriers.
In the field theory limit, a system of coupled pendulums is described the well-known sine-Gordon model.
The introduction of rigid barriers, limiting the motion of the pendulums, makes the potential non-analytical \cite{Arodz:2002yt}.
The limitation of the values of the angle can be dealt with by the use of the so-called ``unfolding'' transformation, yielding a model with potential made up of pieces of the cosine function \cite{Arodz:2005}.
If we further impose that the barriers limit the pendulums to small angles, the cosine function can be approximated by a quadratic function.
In this case we get a displaced and rescaled version of the model in \eqref{eq0}.

Another way of obtaining this potential is through the first BPS submodel of the Skyrme model (BPS---abbreviation from Bogomolny--Prasad--Sommerfield, see \cite{Bogomolny:1975de,Prasad:1975kr}).
It was shown in \cite{Adam:2017pdh} that the Skyrme model can be written combining two BPS submodels.
The angular part of the first of this submodels can be solved through a rational map ansatz, while the radial part is equivalent to a field theory in $1+1$ dimensions, with a limitation to the possible field values.
This limitation can be avoided by an ``unfolding'' transformation, yielding the model with action~\eqref{eq0}, as discussed in \cite{Klimas:2018woi}.

The Euler-Lagrange equation has the following form
\begin{equation}
    \partial_t^2 \eta - \partial_x^2 \eta + V'(\eta) = 0 \label{eq1}
\end{equation}
where  $\eta\neq \eta_\text{min}$  and
\begin{equation}
    V'(\eta) = \sum_{n=-\infty}^\infty \left[ \sgn(\eta - 4n) - (\eta - 4n) \right] H_n(\eta)\label{V'}.
\end{equation}
At the points $\eta=\eta_\text{min}$ the derivative $V'(\eta)$ is not defined (in the classical sense) because the left-hand derivative and the right-hand derivative are not equal at the minima of the potential. This is a very typical situation for models with V-shaped potentials. Looking at the energy
\begin{equation*}
    E=\int dx\left[\frac{1}{2}(\partial_t\eta)^2+\frac{1}{2}(\partial_x\eta)^2+V(\eta)\right]
\end{equation*}
we note that  it is minimized, $E=0$, by static configurations $\eta_\text{min}=2k$, where $V(\eta_\text{min})=0$. Such configurations are physical ones. They obey equation \eqref{eq1} with $V'(\eta)$ replaced by zero. Formally one can define the symbol
\begin{equation}
    [V'](\eta):=
    \begin{cases}
        V'(\eta) & \text{ for } \eta\neq 2k, \\
        0 & \text{ for } \eta=2k.
    \end{cases}
    \label{newV'}
\end{equation}
A similar approach for the signum-Gordon model is given by the extension of the signum function, i.e.\ by assuming that $\sgn(0):=0$. In fact, \eqref{newV'} follows directly from \eqref{V'} with the extension of the signum function. Hence, the final version of the field equation is given by \eqref{eq1} with $V'$ replaced by $[V']$.
The potential and its derivative are shown in Fig.~\ref{fig:VdV}.

There are infinitely many vacuums $\eta = 2k$,  $k \in \mathbb{Z}$ in our model. The  solutions  that interpolate between each two adjacent vacuums are kink (increasing function) and antikink (decreasing function). For instance, the vacuum $\eta = 0$ and $\eta = 2$ are connected by two solutions
\begin{align}
    \eta_{K}(x)      & =
    \begin{cases}
        0               & \text{for } x < a               \\
        1 - \cos(x - a) & \text{for } a \le x \le a + \pi \\
        2               & \text{for } \pi < x
    \end{cases}, \label{eq:kink}\\
    \eta_{\bar K}(x) & =
    \begin{cases}
        2               & \text{for } x < a               \\
        1 + \cos(x - a) & \text{for } a \le x \le a + \pi \\
        0               & \text{for } a + \pi < x
    \end{cases}. \nonumber
\end{align}
Both these solutions are compact, i.e.\ their support is given by $ x \in [a, a + \pi] $.
In general, we can take $a = 0$ and set the kink position by a translation $\eta_K(x - a)$.

The static kink and antikink are Bogomolny solutions, i.e.\ they satisfy two first order differential equations. This can be seen using the Bogomolny trick where one writes the energy density as a total square containing sum or difference of two terms and adding/subtracting the product of these terms. This construction is formalized in \cite{Adam:2013hza}. The applications of this construction to one dimensional scalar field theories is given in \cite{Ferreira:2018ntx} and some further applications to Skyrme model can be found in \cite{Ferreira:2020xcu}. Here we are interested in theories in one spatial dimension. In such approach the topological charge is defined as the integral $Q=\int_{-\infty}^{\infty} dx\, {\cal A}_{\alpha}\widetilde {\cal A}_{\alpha}$ where ${\cal A}_{\alpha}$ and $\widetilde {\cal A}_{\alpha}$ are two functions of the field(s). The index $\alpha$ is formal, its meaning depends on the considered model. The topological character of $Q$ means that it is invariant under smooth infinitesimal field variations, $\delta Q=0$. It leads to an identity relation that involves ${\cal A}_{\alpha}$ and $\widetilde {\cal A}_{\alpha}$. Imposing the self-duality  equations ${\cal A}_{\alpha}=\pm \widetilde {\cal A}_{\alpha}$ one gets from this identity some differential equations that can be interpreted as the Euler-Lagrange equations of a field theoretic model with the energy functional $E=\frac{1}{2}\int_{-\infty}^{\infty} dx\, ({\cal A}_{\alpha}^2+\widetilde {\cal A}_{\alpha}^2)$. Applying the Bogomolny trick one gets $E=\frac{1}{2}\int_{-\infty}^{\infty} dx\, ({\cal A}_{\alpha}\mp \widetilde {\cal A}_{\alpha})^2\pm \int_{-\infty}^{\infty} dx\, {\cal A}_{\alpha}\widetilde {\cal A}_{\alpha} $. The field configuration which satisfies ${\cal A}_{\alpha}=+\widetilde {\cal A}_{\alpha}$ is called self-dual solution whereas that which satisfies ${\cal A}_{\alpha}=- \widetilde {\cal A}_{\alpha}$ is called anti-self-dual solution. The energy of such static field configurations given by $E=\pm Q$.

In our case ${\cal A}_{\alpha}\equiv  \partial_x\eta_{K/\bar K}$ and $\widetilde {\cal A}_{\alpha}\equiv\sqrt{2V(\eta_{K/\bar K})}$ hence the BPS equations read
\begin{equation*}
    \partial_x\eta_{K/\bar K}=\pm\sqrt{2V(\eta_{K/\bar K})}.
\end{equation*}
The self-dual equation corresponds to the plus sign whereas the anti self-dual equation corresponds to the minus sign. The potential evaluated at these solutions, $0\le \eta_{K/\bar K}\le 2$, reads \[V(\eta_{K/\bar K})=\eta_{K/\bar K}-\frac{1}{2}\eta_{K/\bar K}^2=\frac{1}{2}\sin^2(x-a)\] where $\sin(x-a)\ge 0$ at the compacton support. The energy of static solutions reads
\begin{equation*}
    E=\int_{-\infty}^{\infty}dx\left[\frac{1}{2}\eta'^2+V(\eta)\right]
\end{equation*}
where $\eta'\equiv \partial_x\eta$.
In the case of compact solutions only the support $[a,a+\pi]$ is relevant because outside it $\eta'=0$ and $V=0$. One can put the energy integral in the form
\begin{align}
    E & =\int_{a}^{a+\pi}dx\left[\frac{1}{2}\eta'^2_{K/\bar K}+V(\eta_{K/\bar K})\right] \nonumber \\
    & =\frac{1}{2}\int_{a}^{a+\pi}dx\left[\eta'_{K/\bar K}\mp \sqrt{2V(\eta_{K/\bar K})}\right]^2\pm\int_{a}^{a+\pi}dx\;\eta'_{K/\bar K}\sqrt{2V(\eta_{K/\bar K})} \label{expand}
\end{align}
where the last term is topological i.e.\ it can be written as a total derivative. Indeed, defining the pre-potential $U(\eta)$ such that $\sqrt{2V}=\frac{d U}{d \eta}$ one gets $\frac{d\eta}{dx}\frac{dU}{d\eta}=\frac{dU}{dx}$.

\begin{figure}
    \subfigure[]
    {\includegraphics[width=0.45\textwidth]{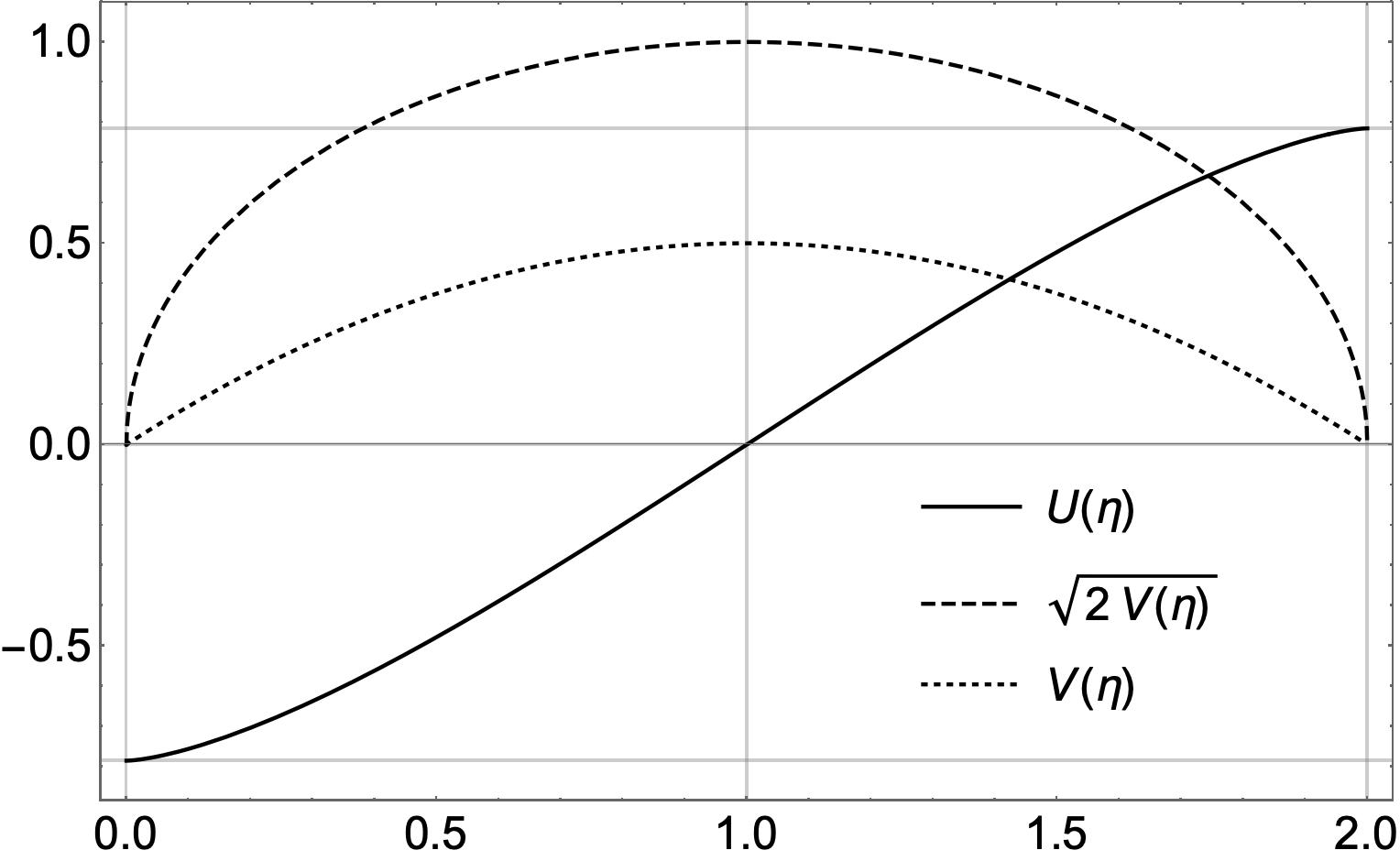}}\hspace{0.5cm}
    \subfigure[]
    {\includegraphics[width=0.45\textwidth]{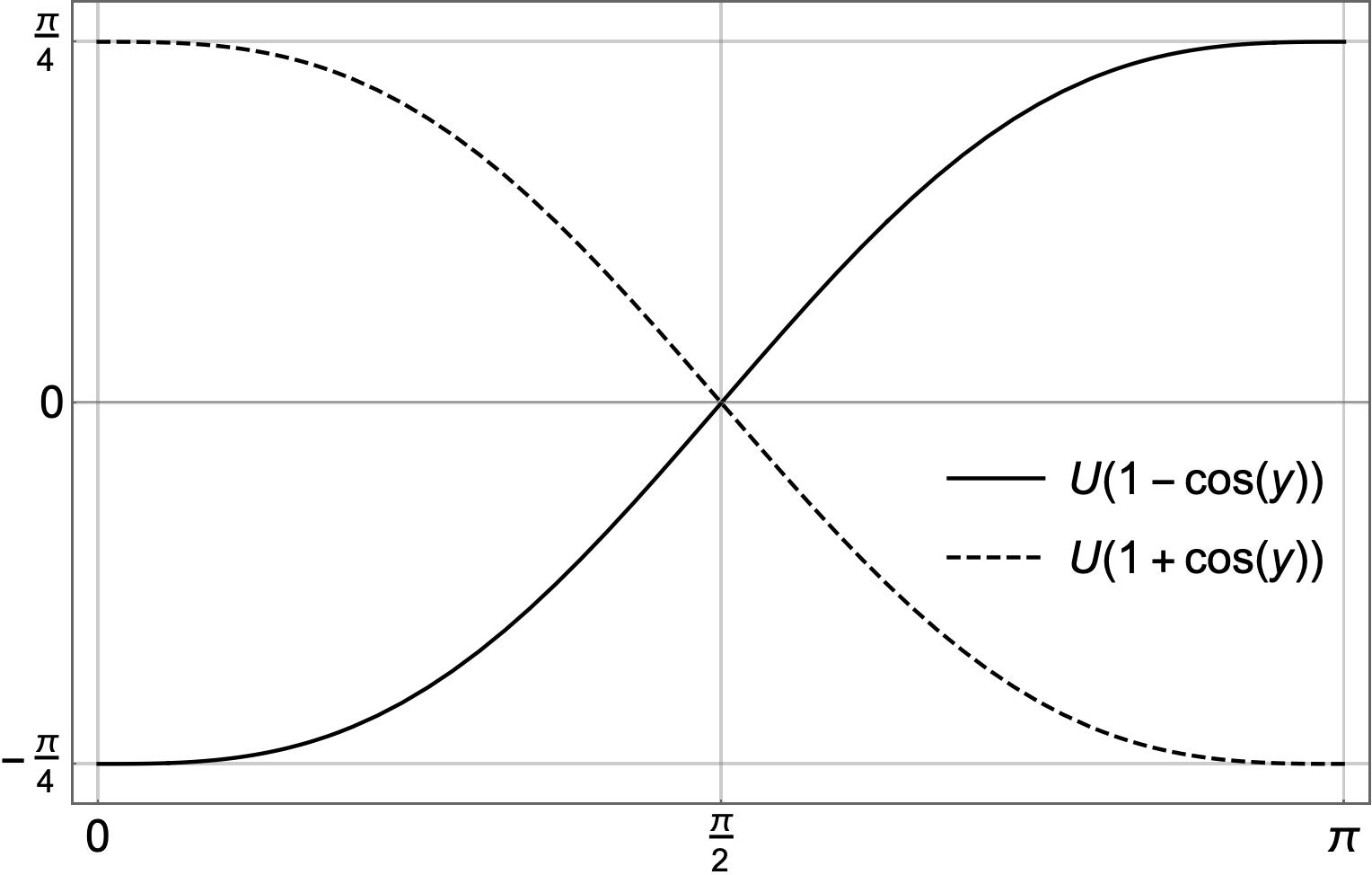}}
    \caption{(a) The pre-potential $U(\eta)$, its derivative $\frac{dU}{d\eta}=\sqrt{2V(\eta)}$ and $V(\eta)=\frac{1}{2}\left(\frac{dU}{d\eta}\right)^2$; (b) Kink $U(\eta_K)$ and antikink $U(\eta_{\bar K})$ in the space of pre-potential in dependence on $y\equiv x-a$.}
    \label{fig:prepotential}
\end{figure}

Since kink and anti kink are the BPS solutions, then the first term vanishes and the energy is equal to the absolute value of the topological charge
\begin{equation*}
    E=\pm \int_{a}^{a+\pi}dx\;\frac{dU}{dx}=\pm\Big(U(\eta_{K/\bar K}(a+\pi))-U(\eta_{K/\bar K}(a))\Big)=\pm Q_\text{Top}.
\end{equation*}
On the other hand, from \eqref{expand} one gets
\[
    E=\pm Q_\text{Top}=\int_{a}^{a+\pi}dx\;\Big(\pm\eta'_{K/\bar K}\Big)\sqrt{2V(\eta_{K/\bar K})}=\int_{a}^{a+\pi}dx\;2V(\eta_{K/\bar K}).
\]
It gives $E=\int_{a}^{a+\pi}dx\;\sin^2(x-a)=\frac{\pi}{2}$. Note that this result does not require the explicit form of the pre-potential. In fact, in this problem the form of the pre-potential can be obtained explicitly, and it reads
\begin{equation*}
    U(\eta)=\frac{1}{2}\left[(\eta-1)\sqrt{2\eta-\eta^2}+\frac{\pi}{2}-\arccos(\eta-1)\right].
\end{equation*}
It follows that $U(\eta_K)=-U(\eta_{\bar K})=\frac{1}{2}\left[-\frac{1}{2}\sin 2(x-a)-\frac{\pi}{2}+x-a\right]$,  which leads to the expression
\[
    U\big(\eta_{K/\bar K}(a+\pi)\big)-U\big(\eta_{K/\bar K}(a)\big)=\pm\frac{\pi}{2}=Q_\text{Top}.
\]
Thus, the energy of static kinks is determined by their topological charge. Note that, in contrary to theories with standard potentials, there are field configurations that consist on arbitrary number of static kinks and antikinks that do not overlap with each other (BPS chains). Such configurations are possible  due to compact nature of kinks.

\section{Shrunken and stretched kinks}
\label{sec:shrunken-and-streched}

To better understand the dynamics of kinks in this model, we will perform a series of numerical simulations. In particular, we are interested in  dynamics  of kinks and their interaction with small field fluctuations which we shall call \emph{radiation}. The appearance of radiation is expected since our model is not integrable. There are many initial field configurations that lead to systems with radiation.
As an example we will consider a system with initial data in the form of perturbed kinks.
This example is useful so that we can see what are the main radiation ingredients in our model.

The simplest kind of perturbation we can apply to a kink is to shrunk or stretch it.
This can be accomplished by a scale transformation in the argument of exact kink solutions.
Therefore, a kink of size $(1 + \epsilon)\pi$ can be obtained by the transformation:
\begin{align}
    \eta(x; \epsilon) &= \eta_K \left( \frac{x}{1 + \epsilon}+\frac{\pi}{2} \right), \label{scale} \\
    \partial_t \eta(x; \epsilon) &= 0 \nonumber
\end{align}
where we perform a translation so that the central point of kink is aligned with the origin.
Such a deformed kink has compact support $x\in [-(1+\epsilon)\pi/2,(1+\epsilon)\pi/2]$.

The dependence of initial configuration on the parameter $\epsilon$ means that energy of this configuration is lower or higher than the BPS kink energy.
The perturbed kink has energy
\begin{equation*}
    E=\int_{-(1+\epsilon)\frac{\pi}{2}}^{(1+\epsilon)\frac{\pi}{2}}dx\Bigg[\frac{1}{2}\eta'^2(x;\epsilon)+V(\eta(x;\epsilon))\Bigg]
    =\left(1+\epsilon+\frac{1}{1+\epsilon}\right)\frac{\pi}{4}.
\end{equation*}
The energy difference $\Delta E=E[\eta(x;\epsilon)]-E[\eta(x;0)]$, reads
\begin{equation*}
    \Delta E= E[\eta(x;\epsilon)]-E[\eta(x;0)]
    = \frac{\epsilon^2}{1+\epsilon}\frac{\pi}{4}.
\end{equation*}
Therefore, such configuration cannot remain static.
We used this perturbed profile as initial conditions for the field equations, and solved it numerically for times $t > 0$.
The numerical work was done using Julia \cite{Julia2017}.
The time integration was performed with the fourth-order Runge-Kutta method and time steps $10^{-4}$ using the library DifferentialEquations.jl \cite{Rackauckas2017}.
The spatial derivative was discretized with steps of $10^{-3}$ with finite differences taken up to the second order.
Other methods, time steps, and discretization schemes were also tried and yielded consistent results.

\begin{figure}
    \includegraphics{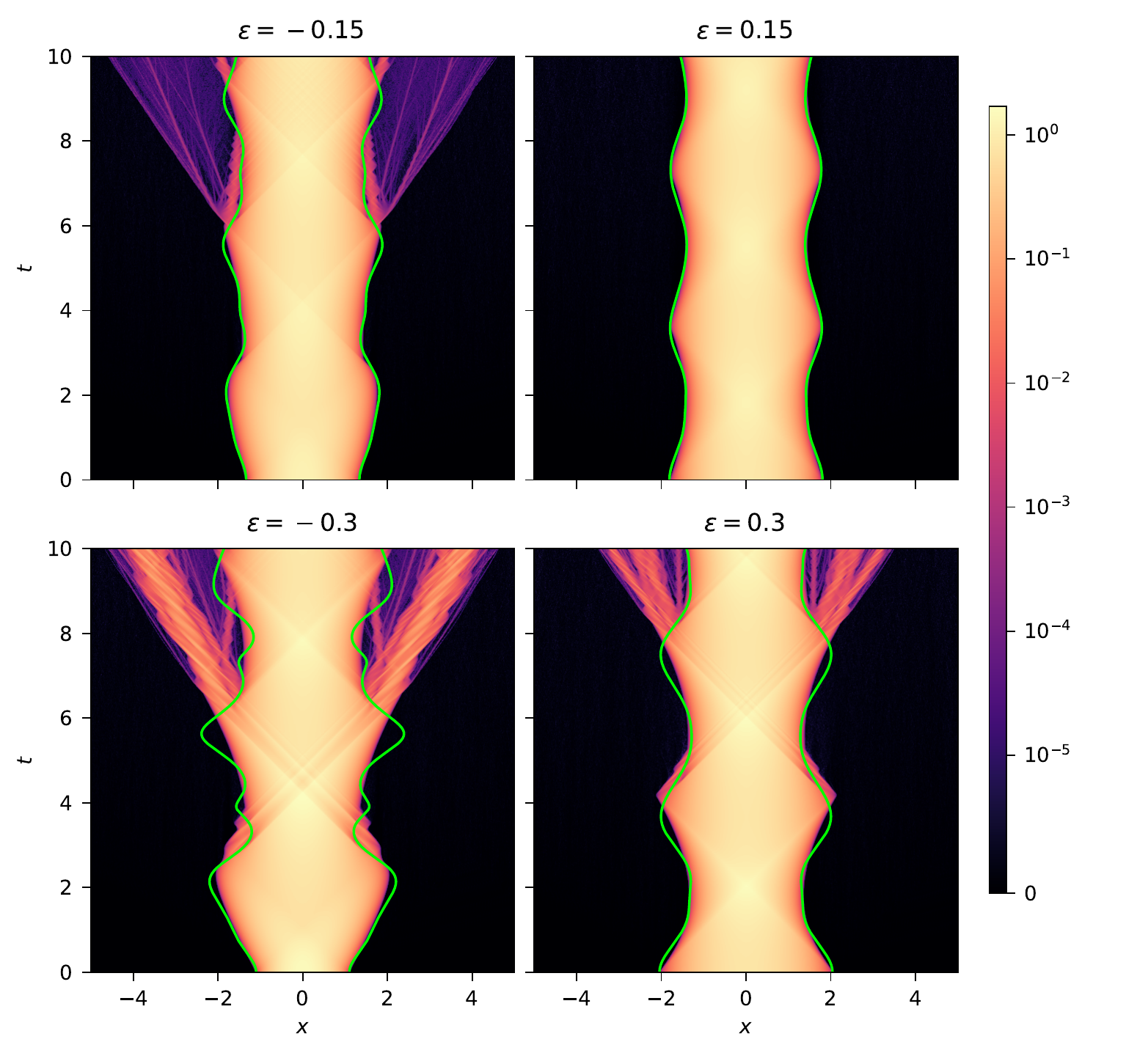}
    \caption{Hamiltonian densities for kinks of size $(1 + \epsilon) \pi $. The green lines show the kink support as predicted from the effective model in equation~\eqref{eq:Leff}.}
    \label{fig:shrunken-and-stretched}
\end{figure}

The Hamiltonian density is presented as a color map gradient on a space-time diagram in Fig.~\ref{fig:shrunken-and-stretched}.
Hamiltonian density plots are used because they allow for easier identification of the radiation than the field plots.
The evolution is such that initially the configuration with $\epsilon>0$ shrinks whereas that with $\epsilon<0$ expands.
The perturbed kinks vibrate in place, releasing energy as radiation.
The radiation is more pronounced in systems with $\epsilon<0$, than in systems with $\epsilon>0$.
This is expected because for the same $|\epsilon|$, the case with $\epsilon<0$ has more excess energy $\Delta E$.

{
It is possible to understand the vibration of the kink support though a collective coordinate approach.
In such approach the field profile is written as a function of $x$ and a finite number of time dependent coordinates, the collective coordinates, effectively reducing the infinitely dimensional field theory to a finite dimensional system.
This is often used in the description of soliton scattering in scalar theories \cite{Sugiyama:1979mi,Manton:2020onl,Manton:2021ipk}.
As a coordinate for this problem we introduce a time dependent scale factor $b(t)$ and write the field profile as a deformation of the static kink in the form
\begin{equation*}
    \eta(t, x) = \eta_K\left( b(t) x + \frac{\pi}{2} \right).
\end{equation*}
The deformation associated with the scale factor $b$, commonly referred as Derrick mode, was previously studied in other models \cite{Rice:1983}, where it is an important ingredient of the dynamics of solitons, for instance in the $\phi^4$ kink-antikink scattering process \cite{Adam:2021gat}.
In terms of this degree of freedom, the action can be rewritten as
\begin{align}
    S &= \int dt \int_{-\pi/2b}^{\pi/2b} dx \, \frac{1}{2} \left( \dot{b}^2 x^2 - b^2 - 1 \right) \sin^2\left(bx + \frac{\pi}{2}\right) \nonumber \\
    &= \int dt \, \frac{\pi}{4\omega^2} \left[ \frac{\dot{b}^2}{b^3} - \omega^2\left(b+\frac{1}{b}\right)\right] \label{eq:b-lag}
\end{align}
where $\dot{b}$ is the derivative of $b$ with respect to time, and $\omega\equiv\sqrt{\frac{12}{\pi^2-6}}$.
This means that the description of the kink vibration can be effectively described by the dynamics of a one-dimensional system with generalized coordinate $b$.
The equation of motion of this system is the Euler-Lagrange equation
\begin{equation*}
    2\frac{ \ddot{b}}{ b^3}-3\frac{\dot{b}^2}{ b^4}+\omega^2\left(1-\frac{1}{b^2}\right)=0.
\end{equation*}
Multiplying this equation by $\dot b$ one gets expression which is a total derivative, and thus it can be integrated giving
\begin{equation*}
    \frac{\dot b^2}{b^3}+\omega^2\left(b+\frac{1}{b}\right)=const.
\end{equation*}
Analyzing the action in \eqref{eq:b-lag}, we can identify the constant as $\frac{4\omega^2}{\pi} E$, where $E$ is the energy.
The second integration gives
\begin{equation*}
    b_\pm(t) = b_0\frac{1 \pm \left(1-b_0^2\right)^{1/2} \left|\sin\left(\omega (t+t_0)\right)\right|}{\cos^2\left(\omega (t+t_0)\right) + b_0^2 \sin^2\left(\omega (t+t_0)\right)}
\end{equation*}
where $b_0$ and $t_0$ are free parameters to be determined from the initial conditions.
These solutions are periodic like the borders of the slightly deformed kinks, however their period $\pi / \omega \approx 1.78$ don't match the numerical results. Note that the function for $b_\pm(t)$ depend on the absolute value of trigonometric function, therefore its period is only half the period of trigonometric functions.

We can improve our results by including extra coordinates to model the kink internal degrees of freedom.
The symmetry of the initial data and the compactness of the solution limit these degrees of freedom to odd functions reaching zero at $x = \pm \pi / 2b(t)$.
Expressing the extras modes as a Fourier series, we write the field configuration as the formula
\begin{equation*}
    \eta(t, x) = \eta_K\left( b(t) x + \frac{\pi}{2} \right) + \sum_{n=1}^{\infty} A_n(t) \sin\left(2 n b(t) x \right)
\end{equation*}
inside its support $x \in [-\pi/2b, \pi / 2b]$.
The continuity of $\partial_x \eta$ at $x = \pm \pi / 2b$ imposes the constraint
\begin{equation}
    \sum_{n=1}^\infty (-1)^n n A_n(t) = 0 \label{eq:constraint}
\end{equation}
to the functions $A_n(t)$.
We can obtain an approximated description by truncating the Fourier series to its first two terms.
The constraint~\eqref{eq:constraint} makes so that the effective degrees of freedom are $b(t)$ and $A_1(t) \equiv A(t)$.
In this case, the field profile is given by
\begin{equation*}
    \eta(t,x) = 1 + \sin (b(t) x) + A(t) \left[\sin (2 b(t) x)+\frac{1}{2} \sin (4 b(t) x)\right].
\end{equation*}
The effective degrees of freedom $b(t)$ and $A(t)$ have its evolution determined by the action
\begin{multline}
    S = \int dt \Bigg[ \left(\left(\frac{\pi ^3}{6}-\frac{275 \pi }{288}\right) A^2+\left(\frac{4 \pi ^2}{15}-\frac{12608}{3375}\right) A+\frac{\pi ^3}{48}-\frac{\pi }{8}\right) \frac{\dot{b}^2}{b^3}
    +\left(\frac{64}{75}-\frac{5\pi  A}{16}\right) \frac{\dot{A} \dot{b}}{b^2}\\
    +\frac{5 \pi  \dot{A}^2}{16 b}
    -\left(2 \pi  A^2+\frac{16 A}{15}+\frac{\pi }{4}\right) b+\left(\frac{5 \pi  A^2}{16}+\frac{16 A}{15}-\frac{\pi }{4}\right)\frac{1}{b}
    \Bigg].
    \label{eq:Leff}
\end{multline}
The equations of motion for $b(t)$ and $A(t)$ are the resulting Euler-Lagrange equations, provided in Appendix~\ref{app:ELbA}.
We solved these equations numerically and presented the resulting kink borders $\pm \pi / 2b(t)$ together with the simulation results in Fig.~\ref{fig:shrunken-and-stretched}.
The interaction between the coordinates $b(t)$ and $A(t)$ allows a more nuanced evolution of the kink borders.
In each case presented, the kink shape evolves differently, and these differences are reflected in the evolution of the collective coordinates.
The agreement between with the numerical results was specially good for the cases with small amounts of radiation, which is the main missing ingredient of our effective description.
In fact, we observe the larger differences in the case $\epsilon=-0.3$ around the times when the deformed kink emit radiation.
}

\begin{figure}
    \includegraphics{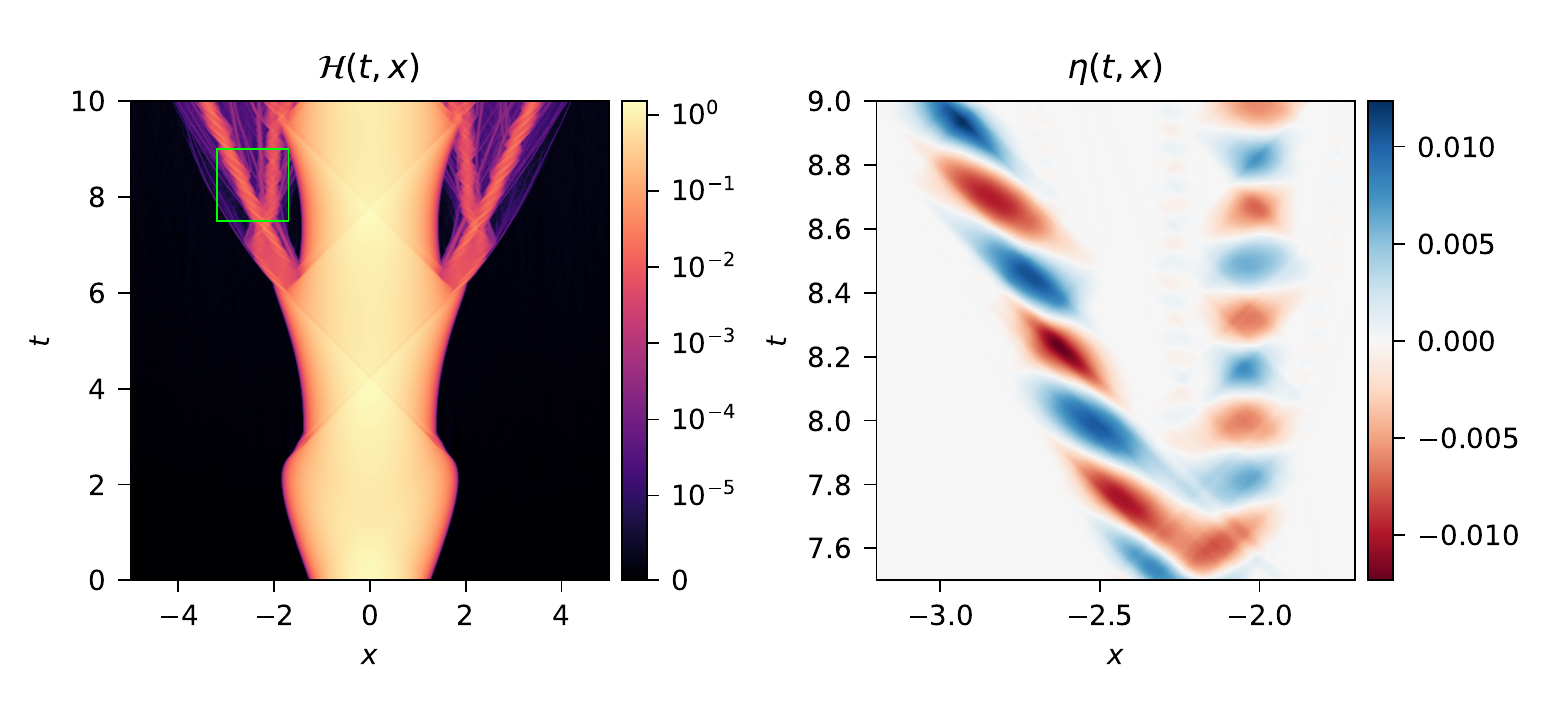}
    \caption{Hamiltonian density (left) and field (right) from simulation of shrunken kink of size $0.8\pi$ ($\epsilon = -0.2$). The field plot is a zoom on the region delimited by the green square on the Hamiltonian density plot.}
    \label{fig:shrunken-and-stretched/zoom}
\end{figure}

It was pointed out in \cite{Klimas:2018woi,Hahne:2019ela,Hahne:2019odw} that the radiation in non-analytic models is dominated by structures similar to oscillons of the signum-Gordon model.
These structures are usually called perturbed or quasi-oscillons for distinction from the exact oscillon solutions.
The presence of these structures is expected here because in the limit of small perturbations around the vacuum, the model \eqref{eq0} can be approximated by the signum-Gordon model.
A closer look at our simulations reveals that shrunken and stretched kinks emit perturbed oscillons, see Fig.~\ref{fig:shrunken-and-stretched/zoom}.
Such oscillons provide an efficient mechanism for emission of surplus energy from larger perturbed oscillons, kinks and other field configurations.
These observations motivated us to study the interaction between kinks and oscillons.

\section{Kink-oscillon scattering}
\label{sec:scattering}

Let us now look at the main problem of the interaction between compact oscillons and compact kinks.
Oscillons are localized, time-dependent and (at least approximately) periodic solutions that are present in some scalar field theories.
Compact and infinitely long-lived oscillons are observed in the signum-Gordon in $1+1$ dimensions \cite{Arodz:2007jh} defined by the action
\begin{equation*}
    S = \int dt \, dx \, \left[ \frac{1}{2}(\partial_t\phi)^2-\frac{1}{2}(\partial_x\phi)^2 - |\phi| \right]
\end{equation*}
and field equation
\begin{equation}
    \partial_t^2 \phi - \partial_x^2 \phi + \sgn\phi = 0 \label{eq:signum-Gordon}
\end{equation}
where we must impose that $\sgn(0) := 0$ so that vacuum configuration $\phi = 0$ satisfies the field equation.
The signum-Gordon model is related to the model \eqref{eq0} in the limit when $\phi = \eta - 2k \ll 1$ for $k \in \mathbb{Z}$, i.e.\ when the field is close to one of the vacuums.
An interesting property of the signum-Gordon oscillon is its swaying motion, which does not change its energy nor its linear momentum.
Such swaying motion can take any shape described by time-like curves.
Both oscillon borders sway in a synchronized manner, so that the oscillon size $l$ remains constant during its period, which is also equal to $l$.
In this work, we will consider oscillons swaying with constant velocity $v_0$, first presented in \cite{Arodz:2011zm}.
More general solutions can be constructed following the steps outlined in \cite{?wierczy?ski2021}.

Oscillon solutions can be constructed from the partial solutions $\phi_k(t, x)$, where $k \in \{ C$, $L_1$, $L_2$, $L_3$, $R_1$, $R_2$, $R_3 \}$ labels the region of space-time where this solution is valid.
During the first half-period, $t \in [0, l/2]$, each partial solution is given by polynomials $\varphi_k$ with the following expressions
\begin{align*}
    \varphi_C(t, x; v_0) &= \frac{((1+v_0) l - 2x)^2 - 4 ((1 + v_0) l - 2v_0x) t + 4 (2 - v_0^2) t^2}{8 (1 - v_0^2)},\\
    \varphi_{L_1}(t, x; v_0) &= \frac{t^2}{2}- \frac{xt}{1 + v_0},\\
    \varphi_{L_2}(t, x; v_0) &= -\frac{(x - v_0t)^2}{2(1 - v_0^2)},\\
    \varphi_{L_3}(t, x; v_0) &= \frac{1}{2} \left(t - \frac{l}{2} \right) \left(t + \frac{l}{2} + \frac{2x - l}{1 - v_0} \right), \\
	\varphi_{R_i}(t, x; v_0) &= \varphi_{L_i}(t, l - x; -v_0).
\end{align*}
Since each partial solution is valid only in a single region, is convenient to introduce the step functions
\begin{align*}
    \Pi_C(t, x; v_0) &= \theta\left(x + t - \frac{l(1+v_0)}{2}\right) \, \theta(x - t) \, \theta\left(-x + t + \frac{l(1+v)}{2} \right) \, \theta(-x - t + l), \\
    \Pi_{L_1}(t, x; v_0) &= \theta(x - t) \, \theta\left(-x - t + \frac{l(1+v_0)}{2} \right),\\
    \Pi_{L_2}(t, x; v_0) &= \theta(x - v_0t) \, \theta(-x + t)  \, \theta\left(-x - t + \frac{l(1+v_0)}{2}\right),\\
    \Pi_{L_3}(t, x; v_0) &= \theta(-x + t) \, \theta\left(x + t - \frac{l(1+v_0)}{2} \right), \\
    \Pi_{R_i}(t, x; v_0) &= \Pi_{L_i}(t, l - x; -v_0).
\end{align*}
Each $\Pi_k$ is equal to one in the region where $\varphi_k$ is a solution and zero otherwise.

\begin{figure}
    \includegraphics[width=1.0\textwidth,height=0.4\textwidth]{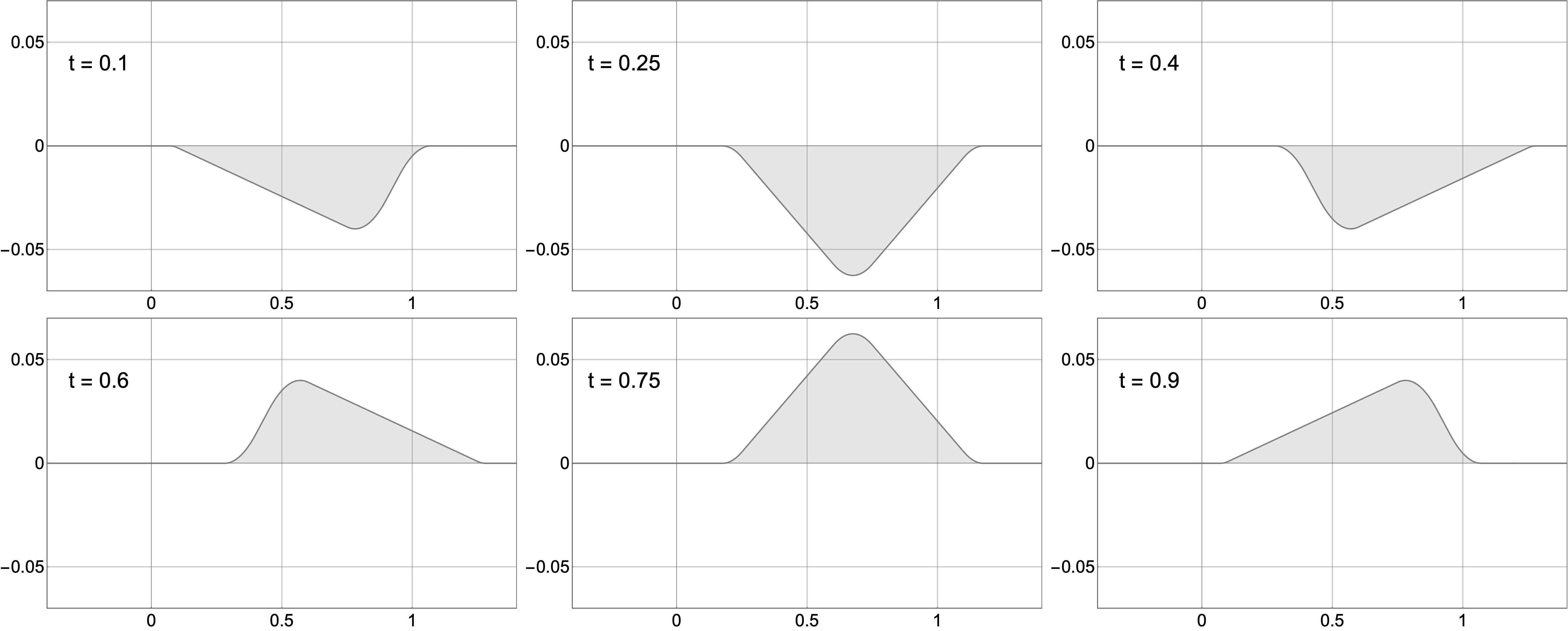}
    \caption{The swaying oscillon with $v_0=0.7$ and $l=1$.}
    \label{fig:swaying}
\end{figure}

Unlike the partial solutions $\phi_k$ that describe the oscillon, the polynomials $\varphi_k$ are not periodic.
Therefore, it is useful to introduce the periodic functions
\begin{align*}
	\tau(t) = \frac{l}{\pi} \arcsin\abs*{\sin \left( \frac{\pi t}{l} \right) }, \qquad
	\sigma(t) = \sgn \left( \sin \left( \frac{2 \pi t}{l} \right) \right)
\end{align*}
such that the function $\tau(t)$ maps the time to the interval $[0, l/2]$, while the function $\sigma(t)$ keeps track of the oscillon sign, which changes each half-period.
Using these functions the partial solutions can be written as
\begin{align*}
    \phi_k(t, x; v_0) = \sigma(t) \, \varphi_k(\tau(t), x; v_0) \, \Pi_k(\tau(t), x; v_0)
\end{align*}
and the complete oscillon solution as
\begin{align*}
    \phi(t, x; v_0) = \sum_k \phi_k(t, x; v_0).
\end{align*}
Such oscillon has energy $E = l^3 / 24$, independent of the swaying parameter $v_0$. In Fig.~\ref{fig:swaying} we show some snapshots of a swaying oscillon with $v_0=0.7$ and $l=1$.

As mentioned before, the oscillon swaying motion is such that its linear momentum is zero, i.e.\ the oscillon center of momentum remains at rest.
However, moving oscillon solutions can be constructed by exploring the Lorentz symmetry of the model.
We write the oscillon profile as $\phi(t', x')$ in the reference frame of its center of momentum.
If such reference frame is moving with velocity $V$ in relation to the laboratory frame, the solution in the laboratory frame is given by
\begin{align*}
    \psi(t, x; V) = \phi(t', x') = \phi\big(\gamma (t - Vx), \gamma(x - V t) \big)
\end{align*}
where $\gamma = (1 - V^2)^{-1/2}$.

\begin{figure}
    \includegraphics{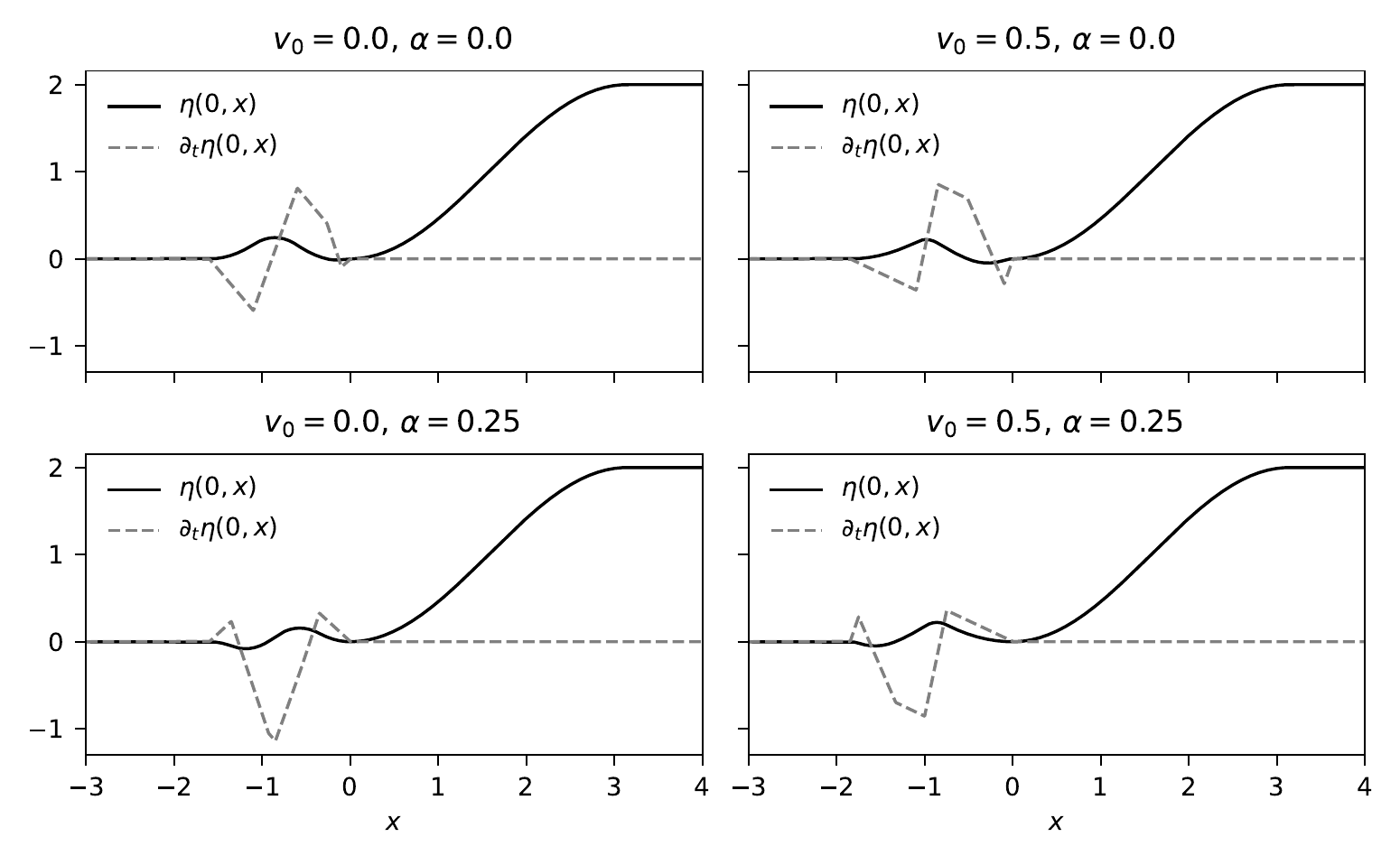}
    \caption{Some initial configurations for a kink-oscillon scattering with parameters $l = 2$, $V = 0.6$, $v_0$, and $\alpha$.}
    \label{fig:kink_oscillon_scattering/initial}
\end{figure}

We will study the interaction between a compact kink and an oscillon trough simulations of scattering processes in the model defined by the action~\eqref{eq0}.
Since both oscillon and kinks have compact support, it is possible to build an initial profile adding together a signum-Gordon oscillon and a kink in a way that their supports do not overlap.
In what follows, we suppose the kink is at rest in the region $x \in [0, \pi]$, and the oscillon is moving towards it from the left.
A uniformly moving oscillon has the following free parameters: rest size $l$, velocity $V$ and swaying motion of its borders.
For simplicity, we will consider only swaying motion with uniform velocity $v_0$ in the oscillon rest frame.
Since signum-Gordon oscillons are only approximated solutions of our model, we must also constrain the oscillon size to be small.
Oscillons are periodic solutions, and we are only interested in the profile for $t = 0$, so there is an additional parameter $\alpha \in [0, 1)$ representing the oscillon phase at the moment of collision.
In terms of a moving oscillon solution from the signum-Gordon model $\psi(t, x; V)$, the oscillon profiles can be written as
\begin{equation*}
    \chi_\text{osc}(t, x) = \psi(t + \alpha l \gamma, x + x_0; V)
\end{equation*}
where the factor of $\gamma = (1-V^2)^{-1/2}$ is needed due to the Lorentzian dilation of the period.
The translation parameter $x_0$ is such that the oscillon right border is located at $x = 0$ when $t=0$.
An explicit procedure for determining $x_0$ can be found in \cite{Hahne:2019ela}.
The full initial conditions are given by
\begin{align*}
    \eta(0, x) & = \eta_K(x) + \chi_\text{osc}(0, x), \\
    \partial_t \eta(0, x) & = \partial_t \chi_\text{osc}(0, x).
\end{align*}
Expressing the initial profile this way, the oscillon-kink collision begins exactly at $t=0$.
Some examples of initial configuration are presented in Fig.~\ref{fig:kink_oscillon_scattering/initial}.

\begin{figure}
    \includegraphics{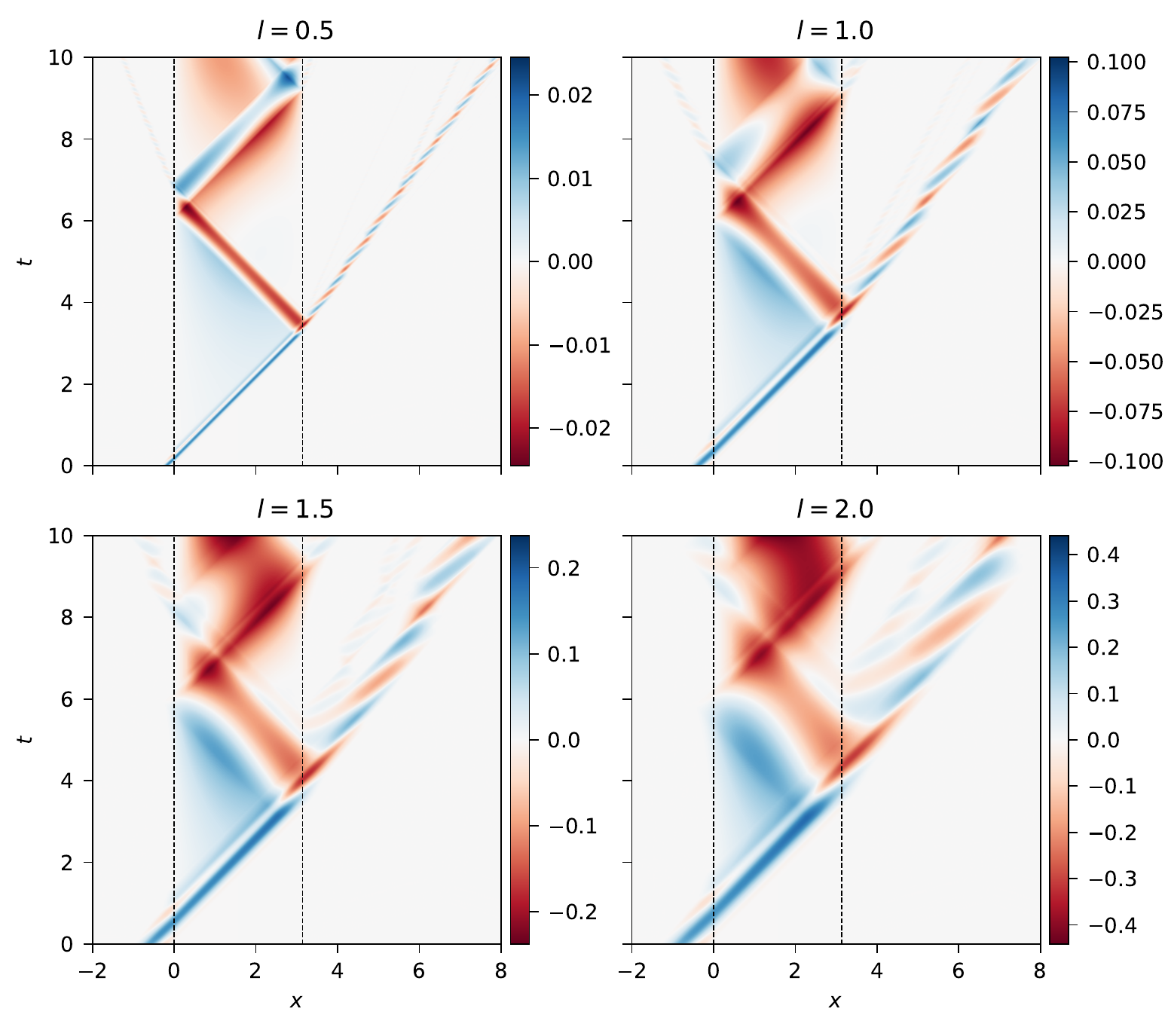}
    \caption{Perturbation $\chi(t,x) = \eta(t,x) - \eta_K(x)$ for parameters $V=0.75$, $\alpha = 0$, $v_0 = 0$, and $l$. The dashed lines show the kink $\eta_K(x)$ support.}
    \label{fig:kink_oscillon_scattering/l}
\end{figure}

\begin{figure}
    \includegraphics[width=\textwidth]{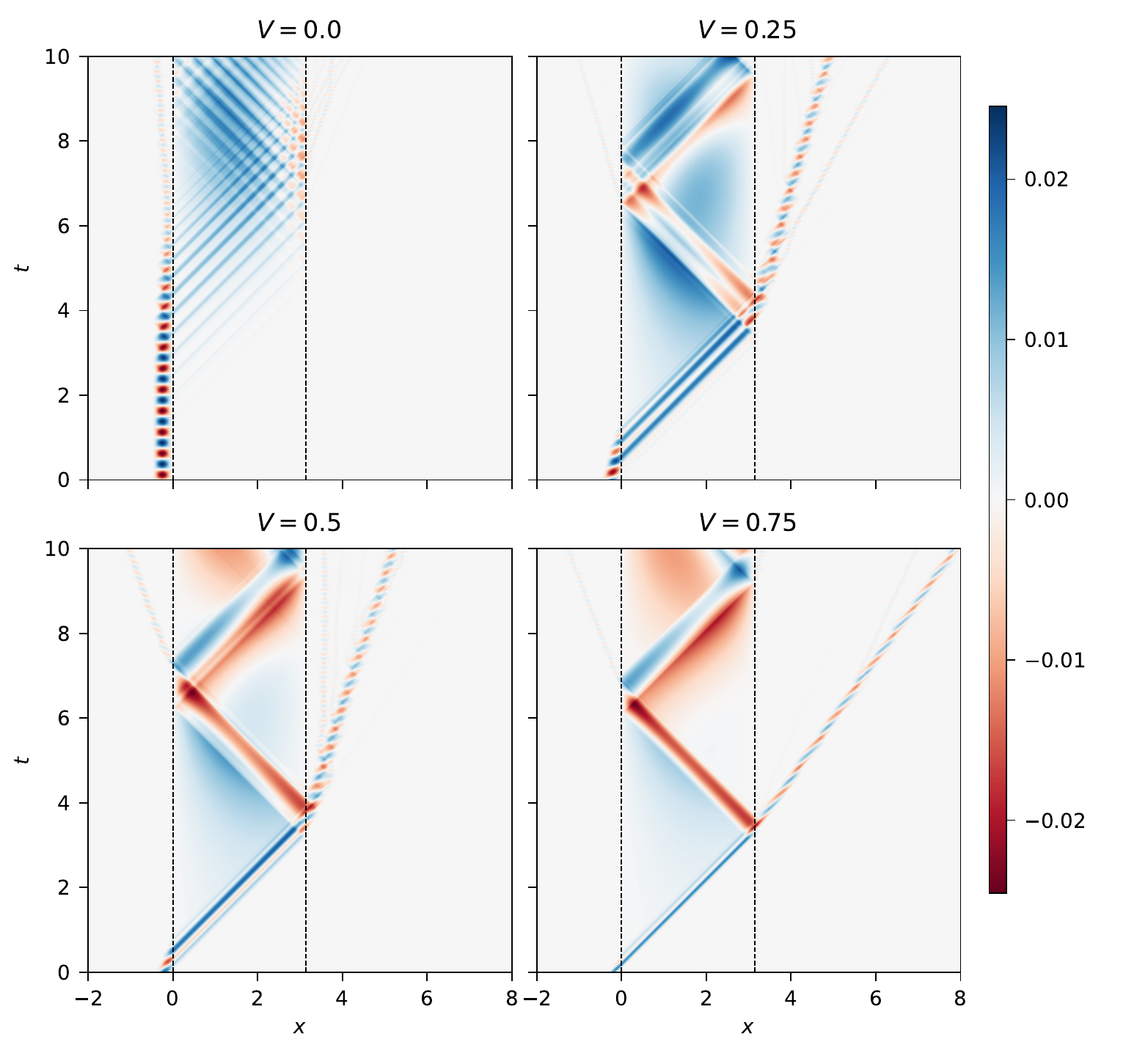}
    \caption{Perturbation $\chi(t,x) = \eta(t,x) - \eta_K(x)$ for parameters $l=0.5$, $\alpha = 0$, $v_0 = 0$, and $V$. The dashed lines show the kink $\eta_K(x)$ support.}
    \label{fig:kink_oscillon_scattering/V}
\end{figure}

\begin{figure}
    \includegraphics{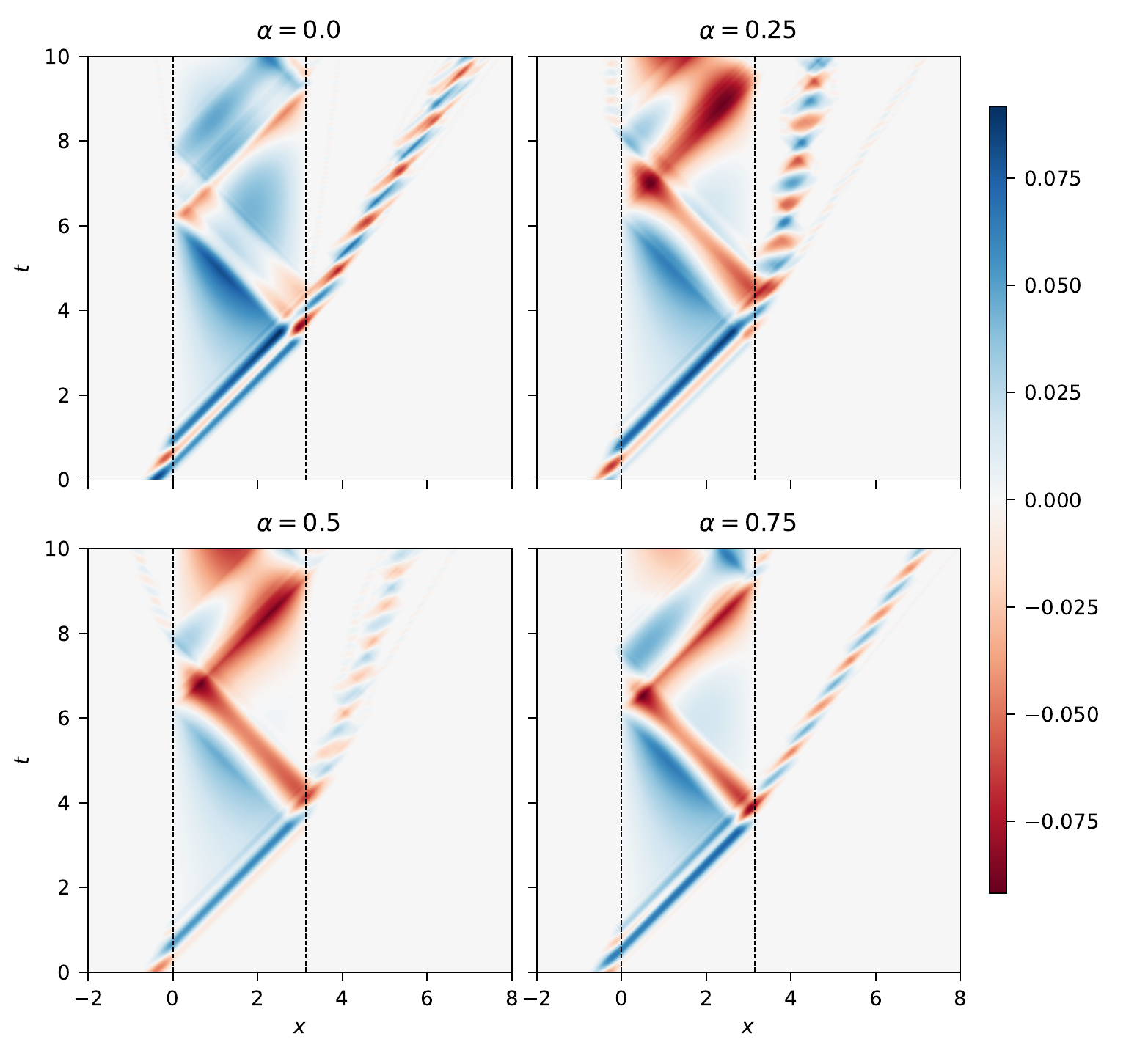}
    \caption{Perturbation $\chi(t,x) = \eta(t,x) - \eta_K(x)$ for parameters $l=1$, $V=0.6$, $v_0 = 0$, and $\alpha$. The dashed lines show the kink $\eta_K(x)$ support.}
    \label{fig:kink_oscillon_scattering/alpha}
\end{figure}

\begin{figure}
    \includegraphics{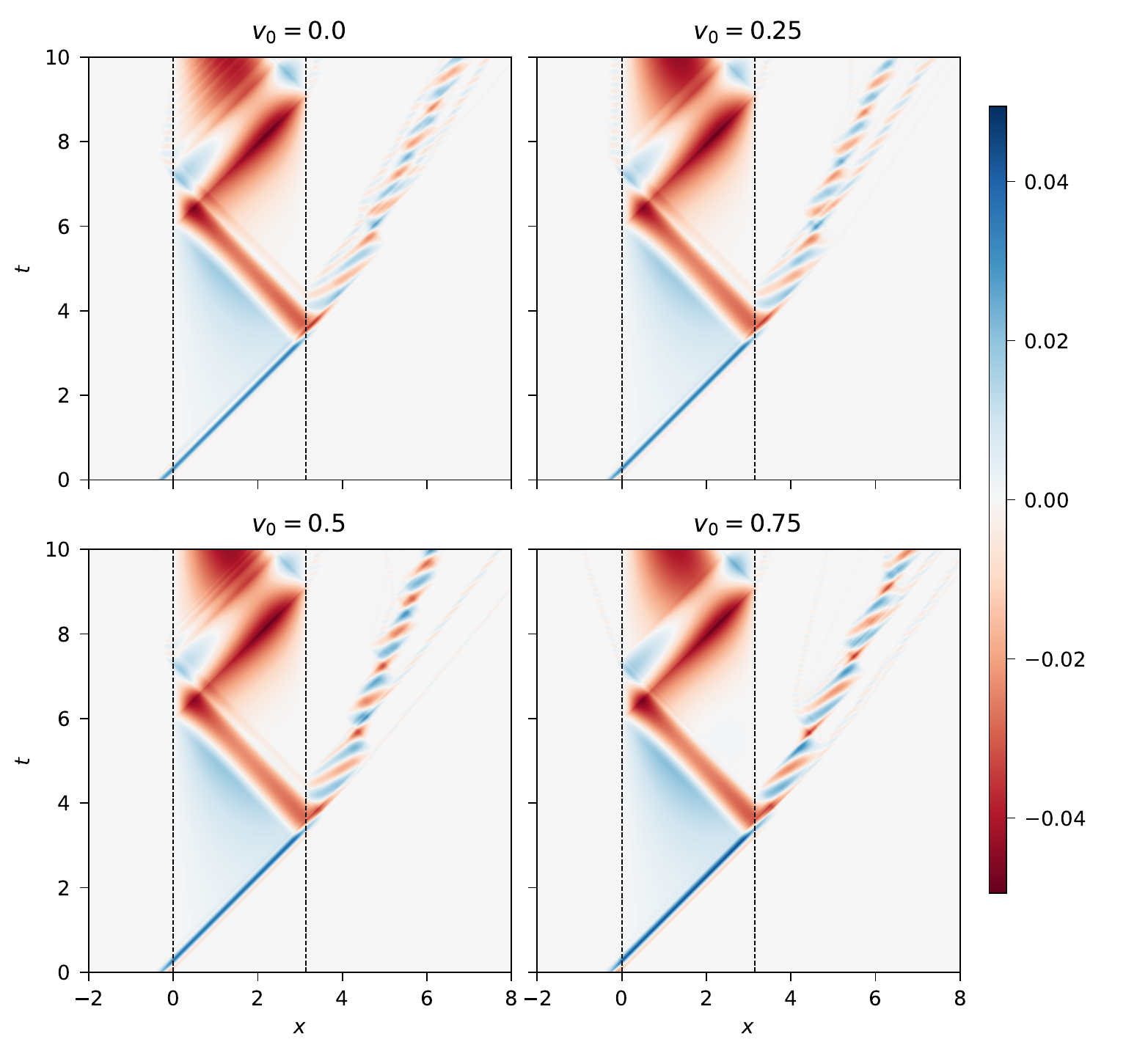}
    \caption{Perturbation $\chi(t,x) = \eta(t,x) - \eta_K(x)$ for parameters $l=0.75$, $V=0.8$, $\alpha = 0$, and $v_0$. The dashed lines show the kink $\eta_K(x)$ support.}
    \label{fig:kink_oscillon_scattering/v0}
\end{figure}

The initial configuration was numerically with the same methods as before.
Since the scattering process did not change the kink location, it is possible to visualize the results through plots of the perturbation alone, defined as the difference
\begin{equation*}
    \chi(t, x) = \eta(t, x) - \eta_K(x).
\end{equation*}
Some simulation results for $\chi(t,x)$ are presented in Fig.~\ref{fig:kink_oscillon_scattering/l}--\ref{fig:kink_oscillon_scattering/v0}.
In each figure one of the simulation parameters was varied, keeping all the others fixed.
This way we can study the dependence from each parameter individually.

Figure~\ref{fig:kink_oscillon_scattering/l} shows simulation results for $V=0.75$, $\alpha=0$, $v_0=0$, and $l$ in the range $0.5$--$2$.
In each case, after the collision the oscillon is completely absorbed by the kink,
creating pulses that travel with the speed of light through the kink bulk.
As the pulse travels, it leaves a trail of perturbation that increases in amplitude with time.
At the kink right border, the pulse changes sign and splits in two.
One part is reflected to the left, carrying the perturbation trail while leaving behind a new one.
Another part is emitted to the right, where it gives origin to perturbed oscillons outside the kink support.
The perturbation is more pronounced for larger values of $l$.
Following the reflected pulse we can see the same process repeating itself at the left border for $l = 0.5$, $1$, and $1.5$.
For $l=2$ there wasn't a new oscillon emission during the simulation.

Figure~\ref{fig:kink_oscillon_scattering/V} shows simulation results for $l=0.5$, $\alpha = 0$, $v_0 = 0$, and $V$ in the range $0$--$0.75$.
The time evolution is strongly dependent on the velocity.
An interesting case is for $V=0$.
If the oscillon dynamics were described by the signum-Gordon equation, its time evolution would be analytically known: it would keep oscillating with period and size $l$ indefinitely.
However, the signum-Gordon oscillon is only an approximated solution to our model.
The behavior of oscillons immersed in the quadratic model was numerically studied in \cite{Klimas:2018woi}.
Small enough oscillons keep oscillating without emitting visible amounts of radiation.
However, its period slowly starts to differ from $l$ and its support slightly vibrates.
These differences are enough to cause oscillon-kink interaction even for $V=v_0=0$.
The interaction causes the emission of pulses through the kink support, visible for $t \gtrsim 3$ when the oscillon has a positive sign.
Since these pulses are very small, emitted oscillons can only be seen for $t \gtrsim 8$.
The original oscillon gradually shrinks and start to move away from the kink.
For velocities $V>0$, the evolution is similar to previous cases.
As $V$ increases, the emitted oscillon is more regular.

Figure~\ref{fig:kink_oscillon_scattering/alpha} shows simulation results for $l=1$, $V=0.6$, $v_0 = 0$, and $\alpha$ in the range $0$--$0.75$.
It is clear that the oscillon phase influences the scattering process.
Primarily, we can see that pulses are more easily created when the oscillon has a positive sign.
The same effect could also be seen in the previous simulations.

Figure~\ref{fig:kink_oscillon_scattering/v0} shows simulation results for $l=0.75$, $V=0.8$, $\alpha = 0$ and $v_0$ in the range $0$--$0.75$.
The influence from the swaying motion parameter $v_0$ is very limited compared to the other parameters.
In each case, the pulse propagates through the kink without any dramatic differences.
However, the emitted oscillon shape still depends on $v_0$, as the trajectories of the smaller radiation components.

\begin{figure}
    \includegraphics{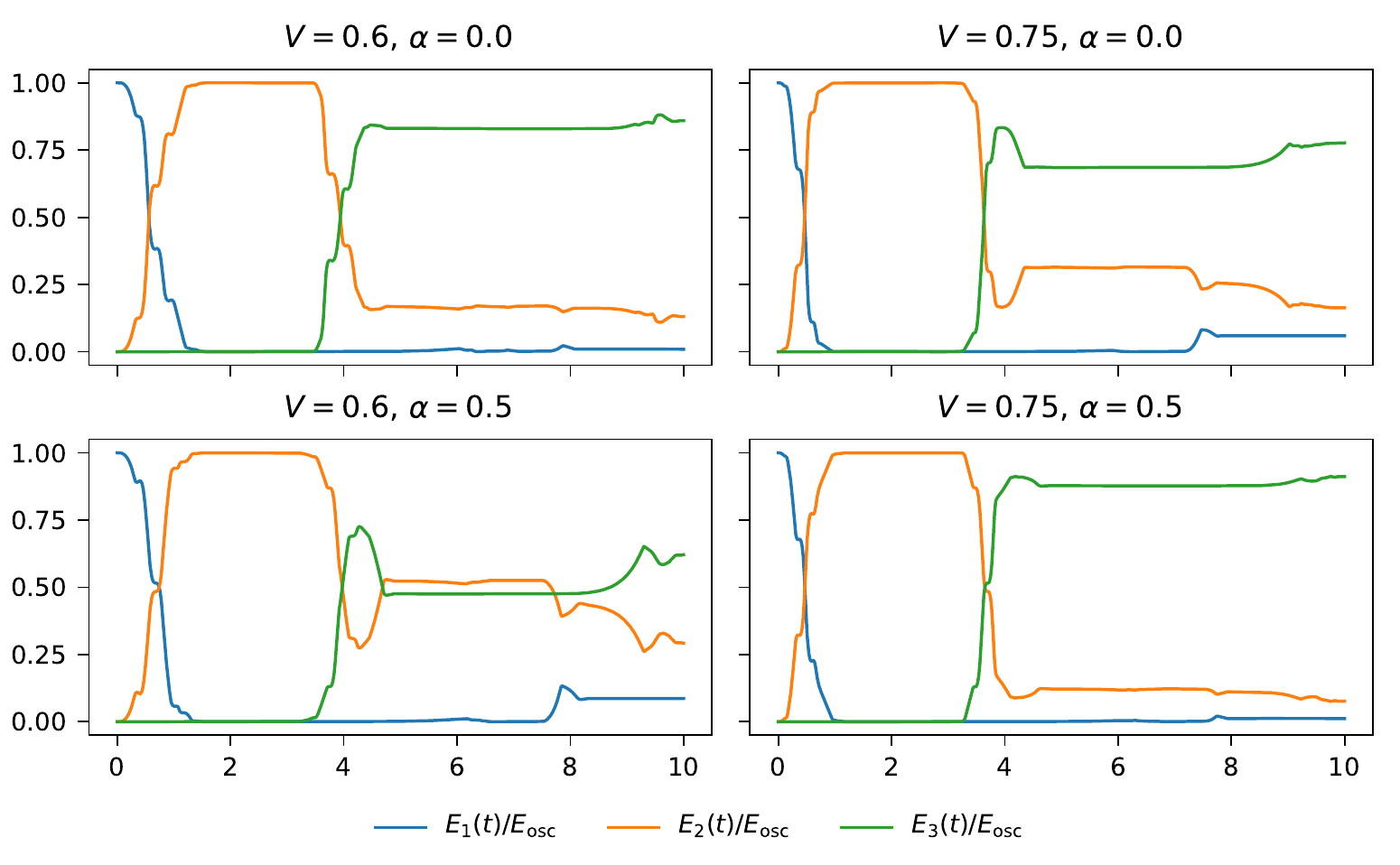}
    \caption{Energies $E_n(t)/E_\text{osc}$ as functions of time for simulations with parameters $l = 1$, $v_0 = 0$, $V$, and $\alpha$.}
    \label{fig:kink_oscillon_scattering/energies-vs-time}
\end{figure}

We can better quantify the dependence from each parameter studying the energy in each region of the $x$ axis.
We define region 1 as the interval $x \in (-\infty, 0]$, region 2 as $x \in (0, \pi]$ and region 3 as $x \in (\pi, \infty)$.
In the perturbation plots, the regions are separated by dashed lines.
In each region, the perturbation energy is given by one of the following expressions:
\begin{align*}
    E_1(t) &= \int_{-\infty}^0 dx \, \left[\frac{1}{2} (\partial_t \eta)^2 + \frac{1}{2} (\partial_x \eta)^2 + V(\eta) \right] \\
    E_2(t) &= \int_{0}^\pi dx \, \left[\frac{1}{2} (\partial_t \eta)^2 + \frac{1}{2} (\partial_x \eta)^2 + V(\eta) \right] - \frac{\pi}{2} \\
    E_3(t) &= \int_\pi^\infty dx \, \left[\frac{1}{2} (\partial_t \eta)^2 + \frac{1}{2} (\partial_x \eta)^2 + V(\eta) \right]
\end{align*}
where we subtracted the kink energy $\pi/2$ from region 2.
These energies are not completely independent of one another, but obey the constraints $E_1(0) = E_\text{osc}$, $E_2(0) = 0$, $E_3(0) = 0$, and $E_1(t) + E_2(t) + E_3(t) = E_\text{osc}$ where $E_\text{osc}$ is the incoming oscillon energy.
Some examples of $E_n(t)$ are shown in Fig~\ref{fig:kink_oscillon_scattering/energies-vs-time}.
In the examples, the oscillon energy is completely transmitted from region 1 to region 2 after the collision.
After the pulses travel through the kink support, an oscillon is emitted in region 3, causing a decrease in $E_2$ and an increase in $E_3$.
However, this emission is not completely efficient, since some energy remains in region 2, corresponding to the reflected pulses.
The efficiency is dependent on the simulation parameters.
To calculate the efficiency systematically, we need to estimate the time $t_1$ for when the first oscillon is completely emitted in region 3.
Such time can be written as
\begin{equation*}
    t_1 = t_\text{a} + \Delta t_\text{t} + \Delta t_\text{e}
\end{equation*}
where $t_\text{a}$ is the time for which the incoming oscillon is absorbed by the kink, $\Delta t_\text{t}$ is the time that takes to the pulse to travel through the kink support, and $\Delta t_\text{e}$ is the time that takes for a complete emission to take place.

The absorption time $t_\text{a}$ can be estimated numerically as the first time for which
\begin{equation}
    E_1(t_\text{a}) \leq \delta \cdot E_\text{osc}. \label{eq:absorption}
\end{equation}
The $\delta$ parameters is the fraction of energy remaining in region 1 after the absorption.
Since the absorption happens quickly, the exact value of $\delta$ is not very important.
In what follows we take $\delta = 0.1$.
The transmission time $\Delta t_\text{t}$ can be estimated by noticing that the pulses travel with the speed of light through a region of size $\pi$.
Therefore, $\Delta t_\text{t} = \pi$.
Finally, the emission time can be estimated through visual inspection of figures \ref{fig:kink_oscillon_scattering/l}--\ref{fig:kink_oscillon_scattering/energies-vs-time}.
Generally the emission takes about a unit of time, therefore $\Delta t_\text{e} = 1$.
Combining all this estimates $t_1 = t_\text{a} + \pi + 1$.
Note that we don't have to be very precise here because $E_n(t)$ plateaus for some time after the oscillon is emitted.

\begin{figure}
    \includegraphics{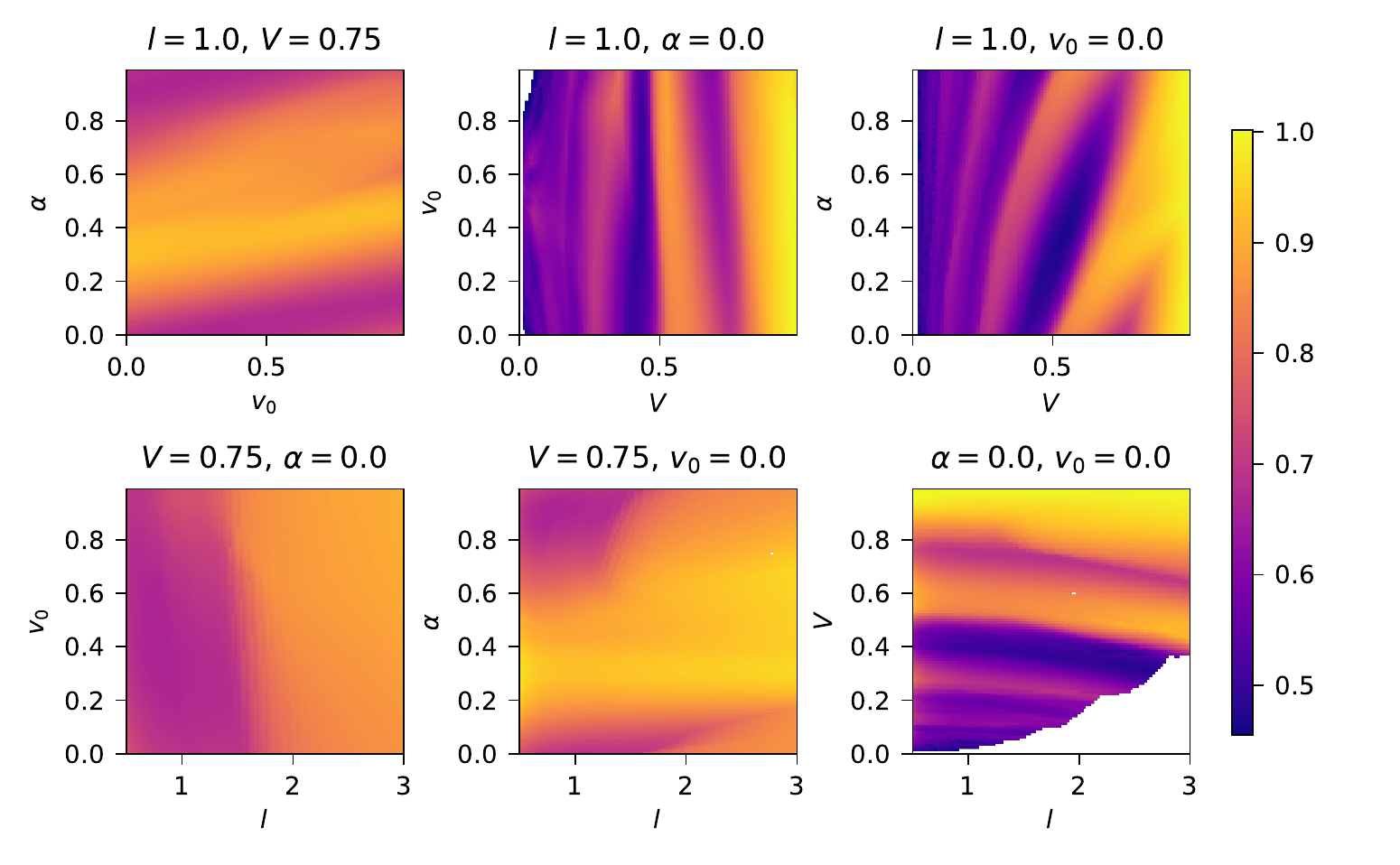}
    \caption{$E_3(t_1)/E_\text{osc}$ as a function of pairs of simulation parameters, keeping fixed the remaining parameters. White regions represent cases for which the condition~\eqref{eq:absorption} was not satisfied for $t_1 \leq 10$.}
    \label{fig:kink_oscillon_scattering/energies}
\end{figure}

Figure~\ref{fig:kink_oscillon_scattering/energies} shows the emitted oscillon energy $E_3(t_1)$ as a fraction of the incoming oscillon energy $E_\text{osc}$.
In each plot, a pair of parameters was fixed while the other two were varied.
Each plot corresponds to about ten thousand simulations with discretization $\Delta x = 5\times 10^{-3}$ and time steps $\Delta t = 5\times 10^{-4}$ for $t \in [0, 10]$.
First thing we notice is that not always the condition~\eqref{eq:absorption} is satisfied.
In particular, for small values of $V$ or large values of $l$, the incoming oscillon was not absorbed by the kink.
Looking at the plots for varying $v_0$ we can confirm the influence of $v_0$ is very small.
The largest dependencies are on the phase $\alpha$ and the velocity $V$.
In particular, although the efficiency usually increases with the velocity, there are ranges of $V$ (determined primarily by the parameters $\alpha$ and $l$) such that the efficiency drops.
This means that $E_3(t)/E_\text{osc}$ depends on $V$ non-monotonically.

\section{Perturbation dynamics}
\label{sec:perturbations}

Most of the scattering dynamics seems to hinge on the propagation of pulses through the oscillon support.
Furthermore, for large velocities, the oscillon profile changes only slightly during the initial instants of the collision.
We will study in details the dynamics of small oscillon-shaped perturbations on top of otherwise exact kinks.
The initial data for this kind of configuration given by
\begin{align*}
    \eta(0, x) &= \eta_{K} (x) + \chi_{\text{osc}} (0, x + x_0), \\
    \partial_{t} \eta(0, x) &= \partial_{t} \chi_{\text{osc}}(0, x + x_0),
\end{align*}
where the translation parameter $x_0$ is chosen such that the oscillon support is covered by the kink support.
A priori, such configuration does not correspond to any physical process containing an oscillon and a kink due to the non-linearity of the field equations.
However, our observations of the scattering process suggest that the perturbation profile travels through the kink support without deforming very dramatically the oscillon shape.
In any case, is useful to look at this kind of configuration to understand the propagation of localized perturbations through a kink.

\begin{figure}
    \includegraphics{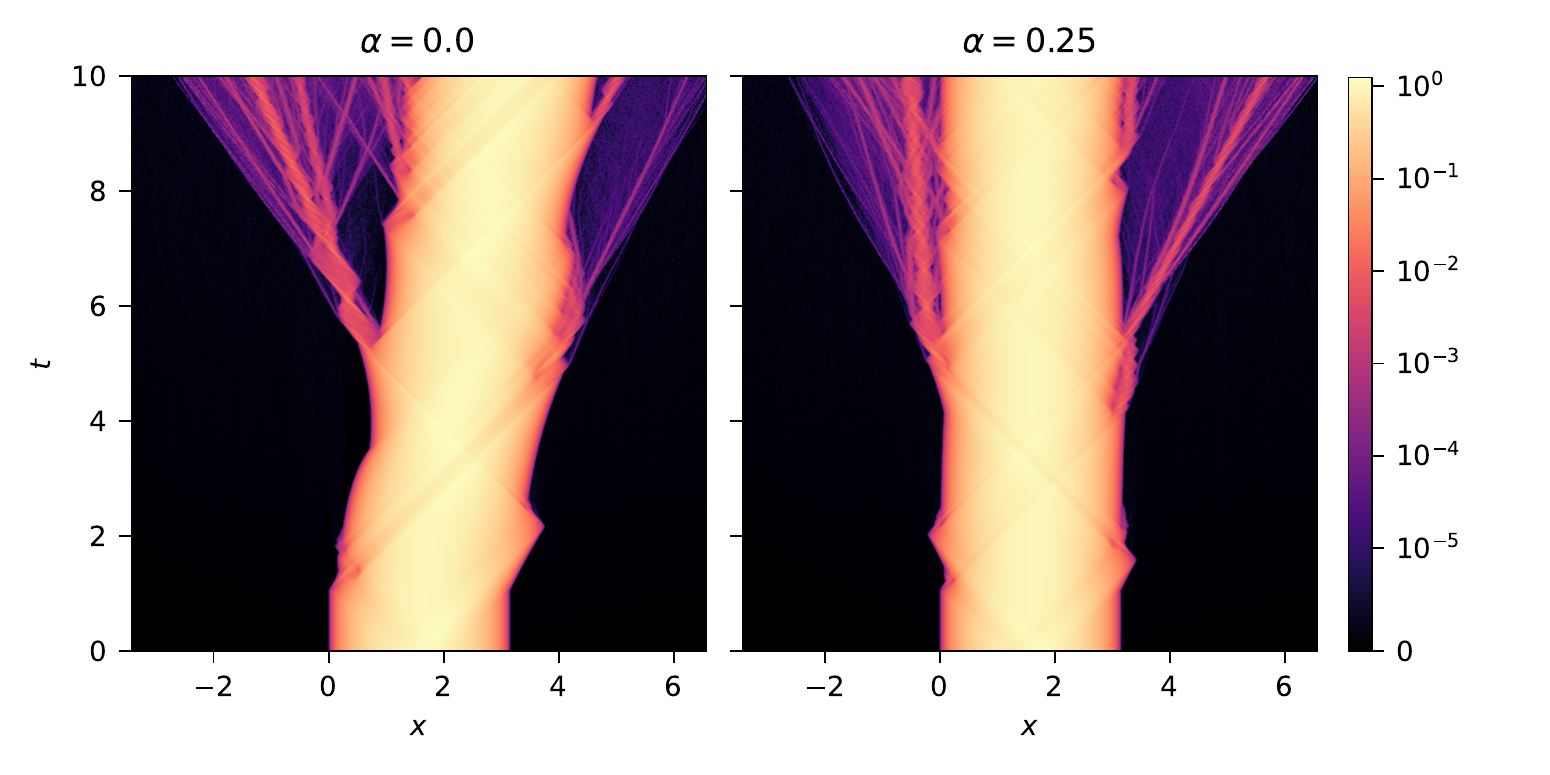}
    \caption{Hamiltonian density for a configuration with an oscillon added on top of a kink. Oscillon and kink have aligned central points. Oscillon parameters are $l=1$, $V=0$, $v_0 = 0$, and $\alpha$.}
    \label{fig:non_zero_p}
\end{figure}

Not all possible configurations reproduce the features of a kink-oscillon scattering.
For example, for $\chi_\text{osc}(t,x)$ representing an oscillon at rest, both the kink and oscillon linear momenta are individually zero.
However, in general the initial profile has non-zero linear momentum
\begin{align*}
    p = -\int_{-\infty}^{\infty} dx \, \partial_t \eta \, \partial_x \eta
    = -\int dx \, \partial_t \chi_{\text{osc}}(0, x) \, \partial_x \eta_{K} (0, x).
\end{align*}
The evolution of this kind of configuration is presented in Fig.~\ref{fig:non_zero_p}.
For $\alpha = 0$, the kink-oscillon superposition has $p \neq 0$, making the kink move to the right.

\begin{figure}
    \includegraphics{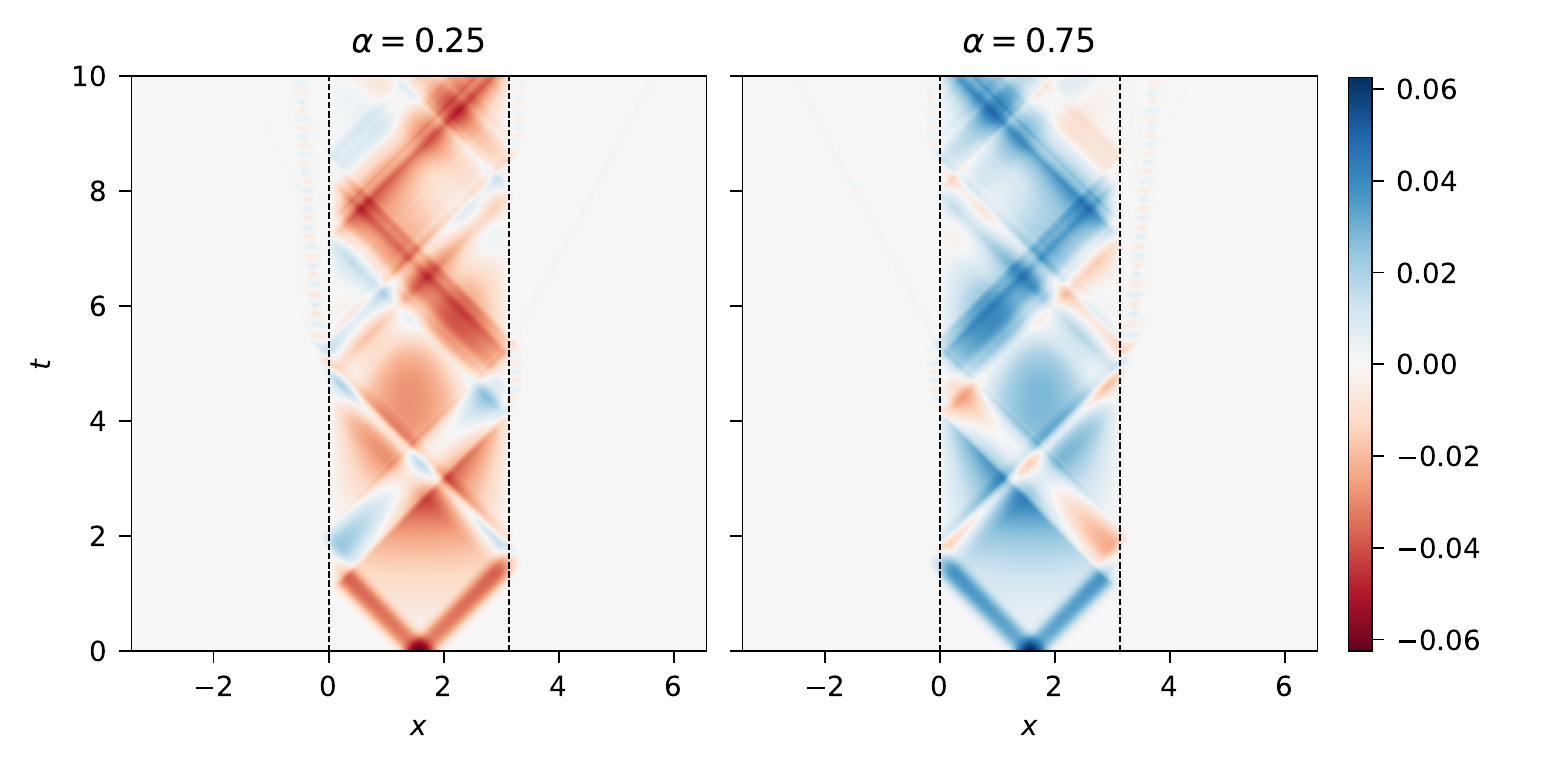}
    \caption{Perturbation $\chi(t, x) = \eta(t, x) - \eta_K(x)$ for a configuration with an oscillon added on top of a kink. Oscillon and kink have aligned central points. Oscillon parameters are $l=1$, $V=0$, $v_0 = 0$, and $\alpha$. Dashed lines indicate the kink $\eta_K(x)$ support.}
    \label{fig:zero_p}
\end{figure}

However, for $\alpha=0.25$, the overall momentum is zero, because $\partial_t \chi_\text{osc}(0.25 l, x) = 0$.
The same is true for $\alpha = 0.75$.
For these cases the kink remains at rest, and it's possible to use the perturbation $\chi(t, x) = \eta(t, x) - \eta_K(x)$ to visualize the results, presented in Fig.~\ref{fig:zero_p}.
Both cases are essentially the same because both the field equations and the initial conditions are symmetric under a reflection around the kink middle point combined with the transformation $\eta \to 2 - \eta$.
In both cases, the oscillon profile creates two pulses propagating with the speed of light at opposite direction.
When the pulses meet the kink borders they are reflected and change sign.
For $t \approx 5$, one pulse gives origin to a perturbed oscillon outside the kink support.

\begin{figure}
    \includegraphics{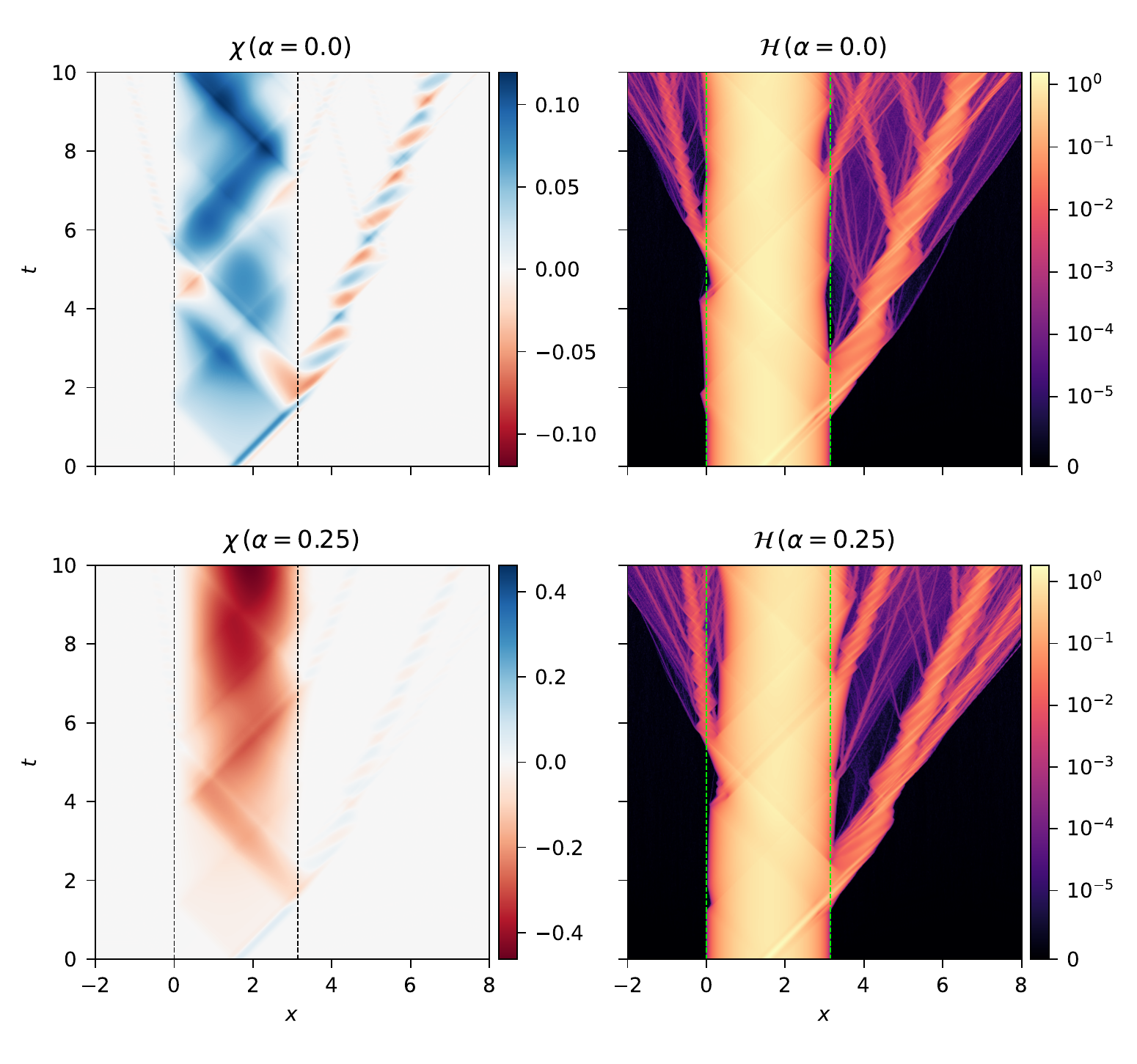}
    \caption{Perturbation $\chi(t,x)$ and Hamiltonian density $\mathcal{H}(t,x)$ for a configuration with an oscillon added on top of a kink. Oscillon and kink have aligned central points. Oscillon parameters are $l=1$, $V=0.75$, $v_0 = 0$, and $\alpha$. For $\alpha=0.25$ the kink starts moving causing $\chi(t,x)$ to grow.}
    \label{fig:kink_oscillon_superposition/centered-alpha}
\end{figure}

\begin{figure}
    \includegraphics{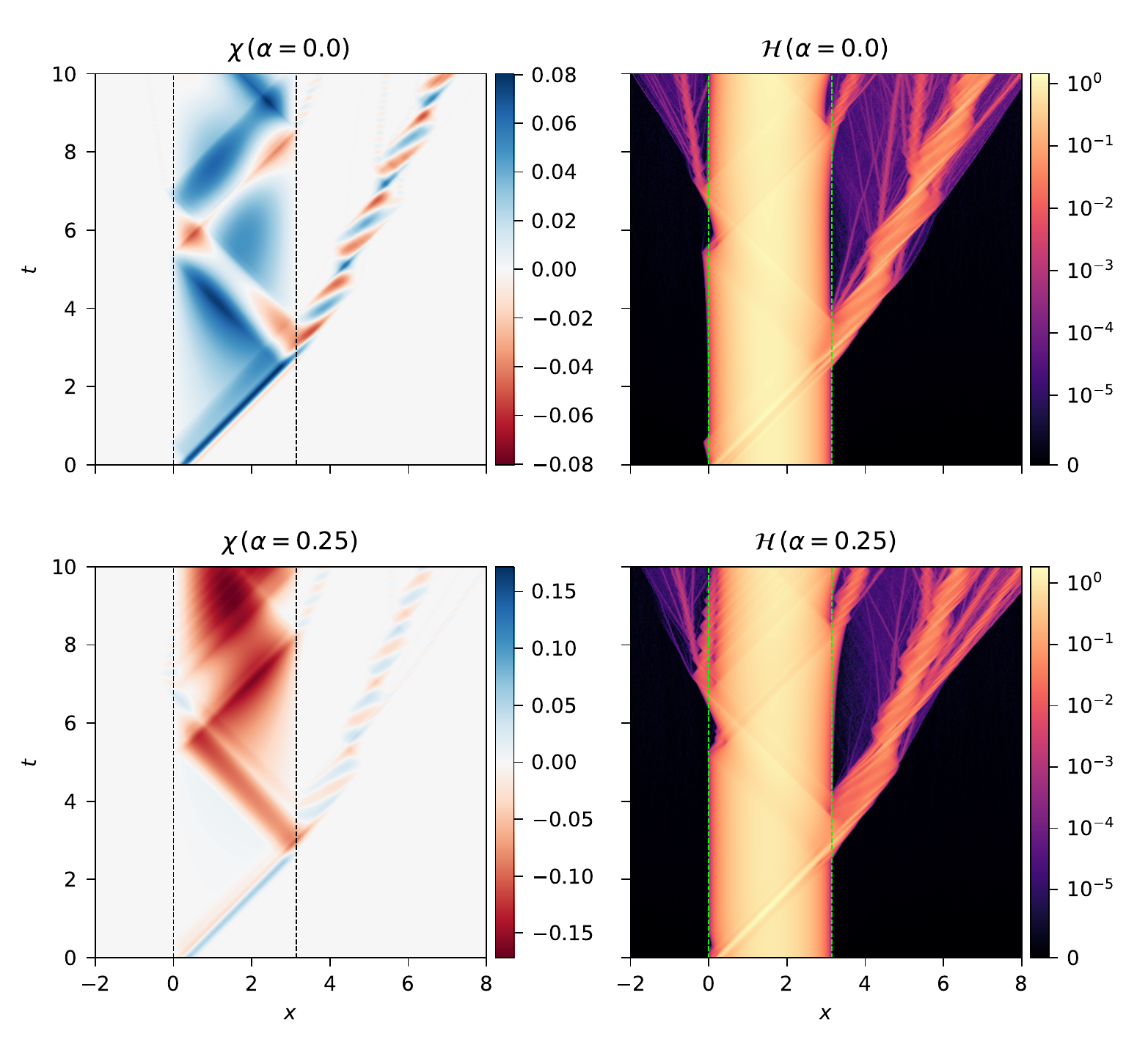}
    \caption{Perturbation $\chi(t,x)$ and Hamiltonian density $\mathcal{H}(t,x)$ for a configuration with an oscillon added on top of a kink. Oscillon and kink have aligned left borders. Oscillon parameters are $l=1$, $V=0.75$, $v_0 = 0$, and $\alpha$.}
    \label{fig:kink_oscillon_superposition/left-alpha}
\end{figure}

Such behavior is very similar to the scattering process.
The main differences are the number of pulses and how easily oscillons are created outside the kink support.
A closer configuration to the scattering process can be considered choosing $\chi_\text{osc}$ to represent an oscillon with $V = 0.75$.
In this case (Fig.~\ref{fig:kink_oscillon_superposition/centered-alpha}) just one pulse is emitted, leaving a perturbation trail behind.
However, for $\alpha = 0.25$, the kink support starts moving, differently from what is observed in the scattering process.
This difference vanishes when the oscillon profile is translated so that kink and oscillon have aligned left borders (Fig.~\ref{fig:kink_oscillon_superposition/left-alpha}).

Under certain conditions, it is possible to obtain the perturbation evolution independently of the simulations.
For values of the field in the range $ -2 < \eta < 2$, the field equation simplifies to
\begin{equation*}
    \partial_t^2 \eta - \partial_x^2 \eta + \sgn(\eta) - \eta = 0.
\end{equation*}
Expressing the field equation in terms of the perturbation $\chi(t, x) = \eta(t,x) - \eta_K(t,x)$ we get
\begin{equation*}
    \partial_t^2 \eta_K - \partial_x^2 \eta_K + \sgn(\eta_K + \chi) - \eta_K + \partial_t^2 \chi - \partial_x^2 \chi - \chi = 0.
\end{equation*}
For perturbations small enough such that $\sgn(\eta_K + \chi) = \sgn\eta_K$, the equation simplifies to
\begin{equation*}
    \partial_t^2 \chi - \partial_x^2 \chi - \chi = 0.
\end{equation*}
This is a Klein-Gordon equation for $m^2 = - 1$.
Its solutions are plane waves $e^{i(kx + \omega_k t)}$ with dispersion relation $-\omega_k^2 + k^2 - 1 = 0$.

Using the dispersion relation, it is possible to write a general expression for the perturbation
\begin{equation}
    \begin{split}
        \chi(t, x) = \int_{k^2 \geq 1} dk \left[ a(k) e^{i \sqrt{k^2 - 1}t} + b(k) e^{- i \sqrt{k^2 - 1}t} \right] e^{i kx} \\
        + \int_{k^2 \leq 1} dk \left[ a(k) e^{\sqrt{1 - k^2}t} + b(k) e^{- \sqrt{1 - k^2}t} \right] e^{i kx}.
    \end{split}\label{eq:chi-ab}
\end{equation}
For initial conditions $\chi(0, x)$ e $\partial_t \chi(0, x)$ known in the form
\begin{align*}
    \chi(0, x) &= f(x) = \int dk \, \tilde{f}(k) \, e^{i kx}, \\
    \partial_t \chi(0, x) &= g(x) = \int dk \, \tilde{g}(k) \, e^{i kx}
\end{align*}
is possible to obtain explicit expressions for $a(k)$ and $b(k)$:
\begin{align*}
    a(k) &=
    \begin{cases}
        \frac{1}{2} \left[\tilde{f}(k) + \frac{1}{i \sqrt{k^2 - 1}} \tilde{g}(k) \right] &\text{ for } k^2 \geq 1, \\
        \frac{1}{2} \left[\tilde{f}(k) + \frac{1}{\sqrt{1 - k^2}} \tilde{g}(k) \right] &\text{ for } k^2 < 1,
    \end{cases} \\
    b(k) &=
    \begin{cases}
        \frac{1}{2} \left[\tilde{f}(k) - \frac{1}{i \sqrt{k^2 - 1}} \tilde{g}(k) \right] &\text{ for } k^2 \geq 1, \\
        \frac{1}{2} \left[\tilde{f}(k) - \frac{1}{\sqrt{1 - k^2}} \tilde{g}(k) \right] &\text{ for } k^2 < 1,
    \end{cases}
\end{align*}
and rewrite the equation~\eqref{eq:chi-ab} as
\begin{equation}
    \begin{split}
        \chi(t, x) = \int_{k^2 \geq 1} dk
        \left[
            \tilde{f}(k) \cos\left( \sqrt{k^2 - 1}t \right)
            + \frac{\tilde{g}(k)}{\sqrt{k^2 - 1}} \sin\left( \sqrt{k^2 - 1}t \right)
        \right] e^{i kx} \\
        + \int_{k^2 \leq 1} dk \left[
            \tilde{f}(k) \cosh\left( \sqrt{1 - k^2}t \right)
            + \frac{\tilde{g}(k)}{\sqrt{1 - k^2}} \sinh\left( \sqrt{1 - k^2}t \right)
        \right] e^{i kx}.
    \end{split}\label{eq:chi-fg}
\end{equation}

Following these steps, we reduced the problem of finding the evolution of $\chi(t,x)$ to a problem of Fourier transforms.
Looking at equation~\eqref{eq:chi-fg} we see that the perturbation is the inverse Fourier transform of the function
\begin{equation*}
    \tilde{\chi}(t, k) =
    \begin{cases}
        \tilde{f}(k) \cos\left( \sqrt{k^2 - 1}t \right)
        + \frac{1}{\sqrt{k^2 - 1}} \tilde{g}(k) \sin\left( \sqrt{k^2 - 1}t \right)
        &\text{ for } k^2 \geq 1, \\
        \tilde{f}(k) \cosh\left(\sqrt{1 - k^2}t \right)
        + \frac{1}{\sqrt{1 - k^2}} \tilde{g}(k) \sinh\left( \sqrt{1 - k^2}t \right)
        &\text{ for } k^2 < 1.
    \end{cases}
\end{equation*}
Unfortunately, such Fourier transform is too complicated to be generally invertible.
However, this kind of problem can be easily solved numerically using the library FFTW \cite{FFTW.jl-2005}.

\begin{figure}
    \includegraphics{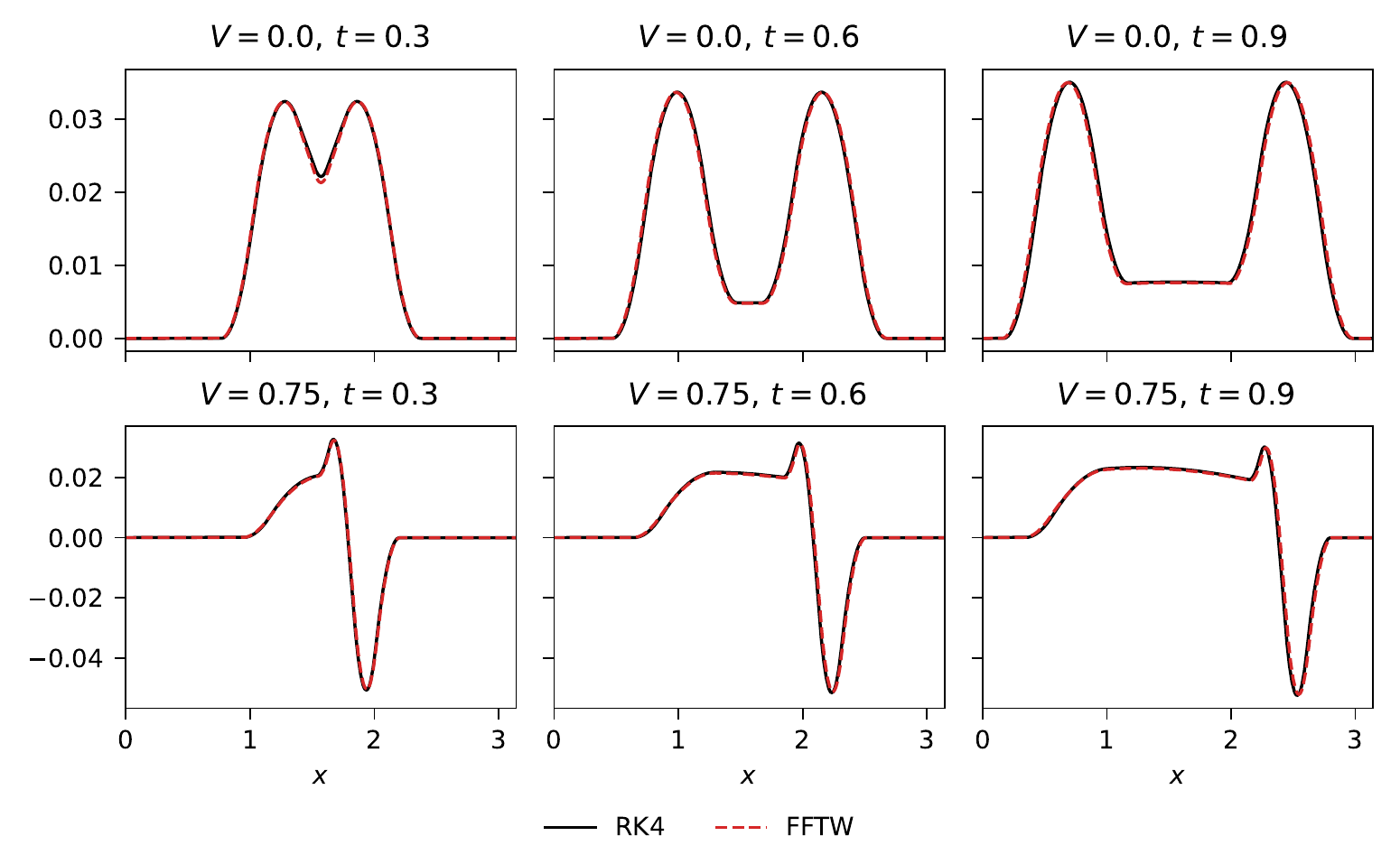}
    \caption{Comparison of values $\chi(t,x)$ obtained through simulation (RK4, black line) and semi-analytical calculation (FFTW, red dashes) for scattering parameter $l = 1$, $\alpha = 0.75$, $v_0 = 0$ and $V$. The first row correspond to $V=0$ and the second row to $V=0.75$. The oscillon was translated so that its central point was aligned with the kink central point.}
    \label{fig:kink_oscillon_superposition/comparison}
\end{figure}

It is important to notice that even using numerical tools again, these are of a fundamentally different nature from the previous simulations.
While the simulations use the complete field equation and initial conditions, the semi-analytical approach delineated here only uses the perturbation initial data.
Therefore, the results from this approach provide us with a way to check the validity of our simulations.
Comparing results from both approaches (Fig.~\ref{fig:kink_oscillon_superposition/comparison}), we can see that they are consistent with each other.

Another important point is that $\chi(t, x)$ has both oscillating modes ($k^2 \geq 1$) and exponential modes ($k^2 < 1$).
If $\chi(t,x)$ represented a free field, the exponential modes would have to be eliminated, otherwise the field would grow indefinitely.
Such elimination would make $\chi(t,x)$ non-causal, allowing for the propagation of faster than light signals.
However, for our purposes, the Klein-Gordon equation is only valid when the condition $\sgn(\eta_K + \chi) = \sgn\eta_K$ is true.
When $\chi(t,x)$ grows large enough to change the sign of $\eta(t,x)$, the solution~\eqref{eq:chi-fg} is no longer valid.
Therefore, we do not have to eliminate the exponential modes and do not have problems with causality \cite{Aharonov1969}.

\section{Toy model}
\label{sec:toy}

\begin{figure}[ht!]
    \includegraphics[width=0.6\textwidth]{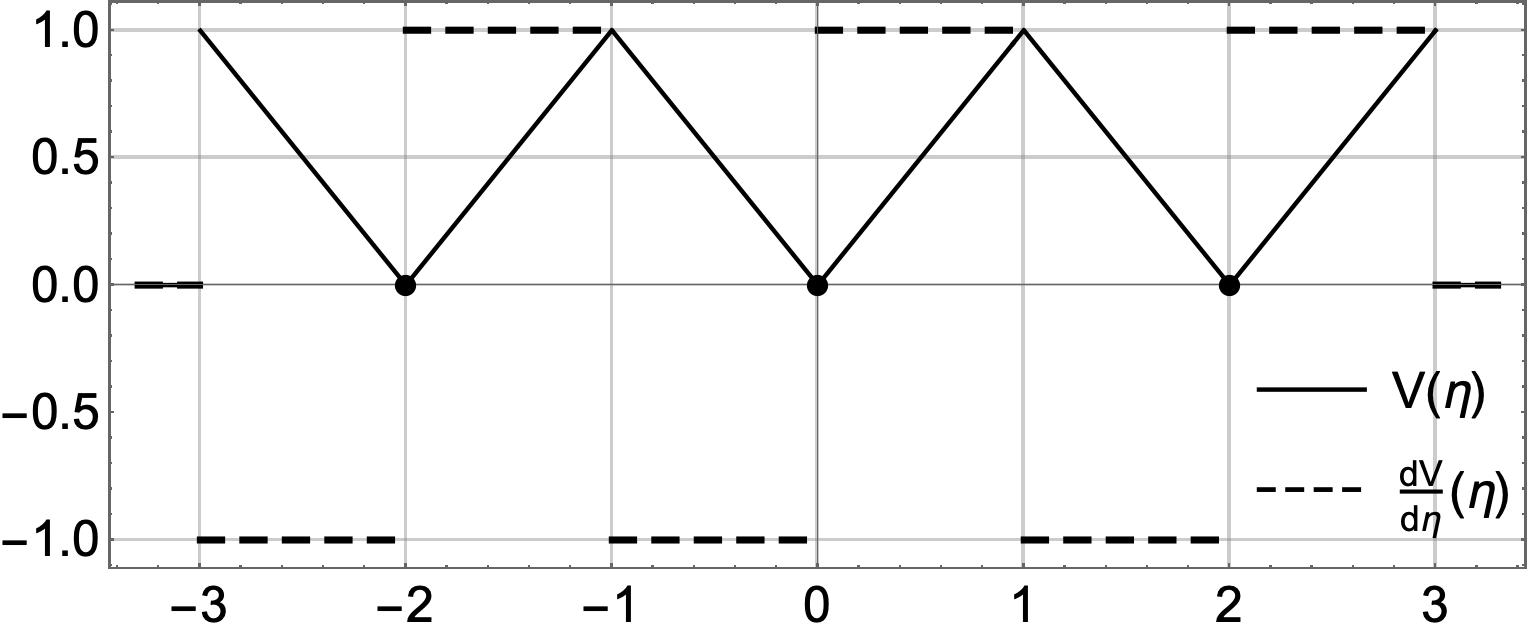}
    \caption{Toy-model potential $V(\eta)$ and its derivative $V'(\eta)$.}
    \label{fig:saw}
\end{figure}

\begin{figure}[ht!]
    \subfigure
    []
    {\includegraphics[width=0.45\textwidth,height=0.25\textwidth]{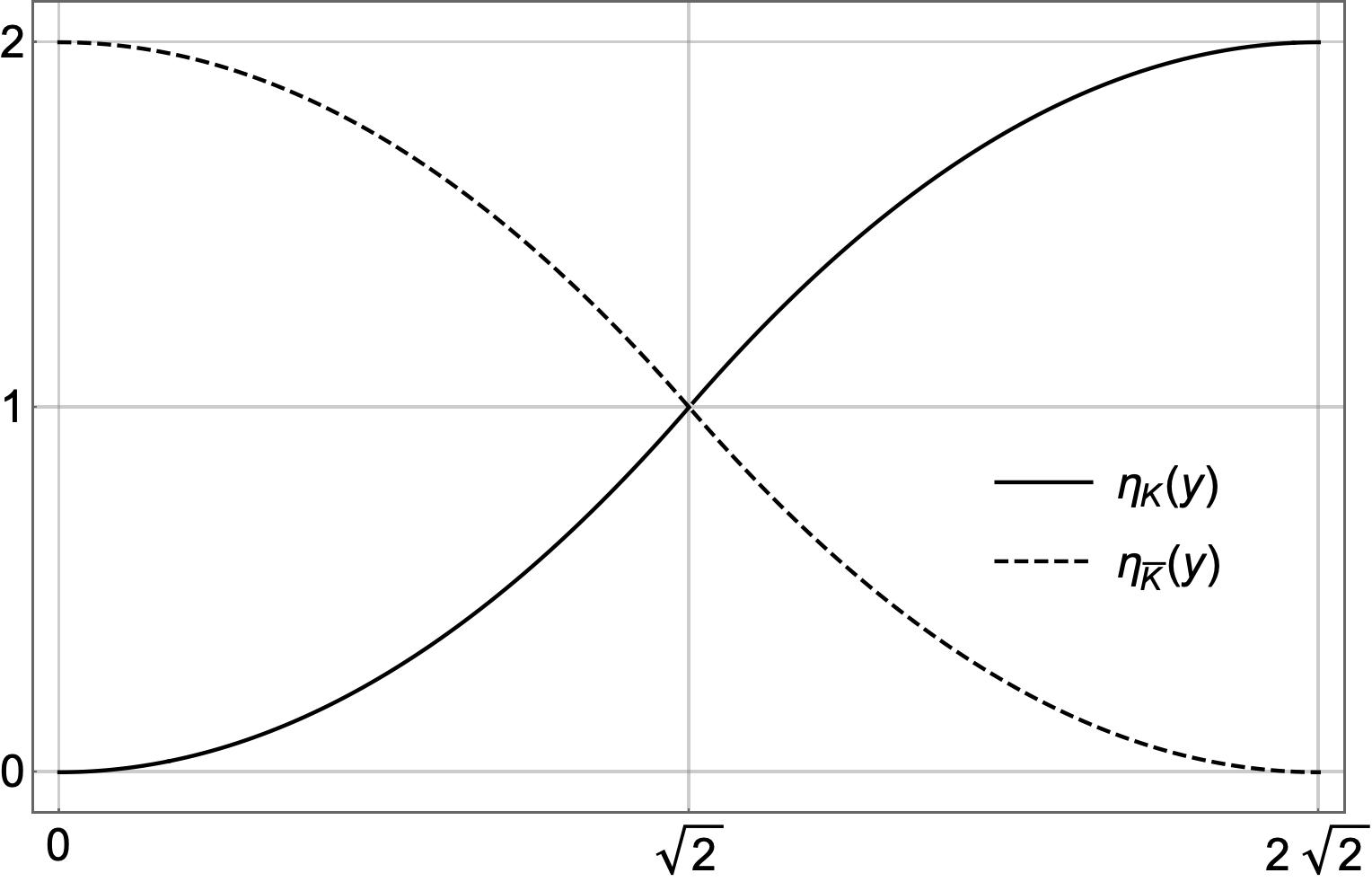}}
    \subfigure
    []
    {\includegraphics[width=0.45\textwidth,height=0.25\textwidth]{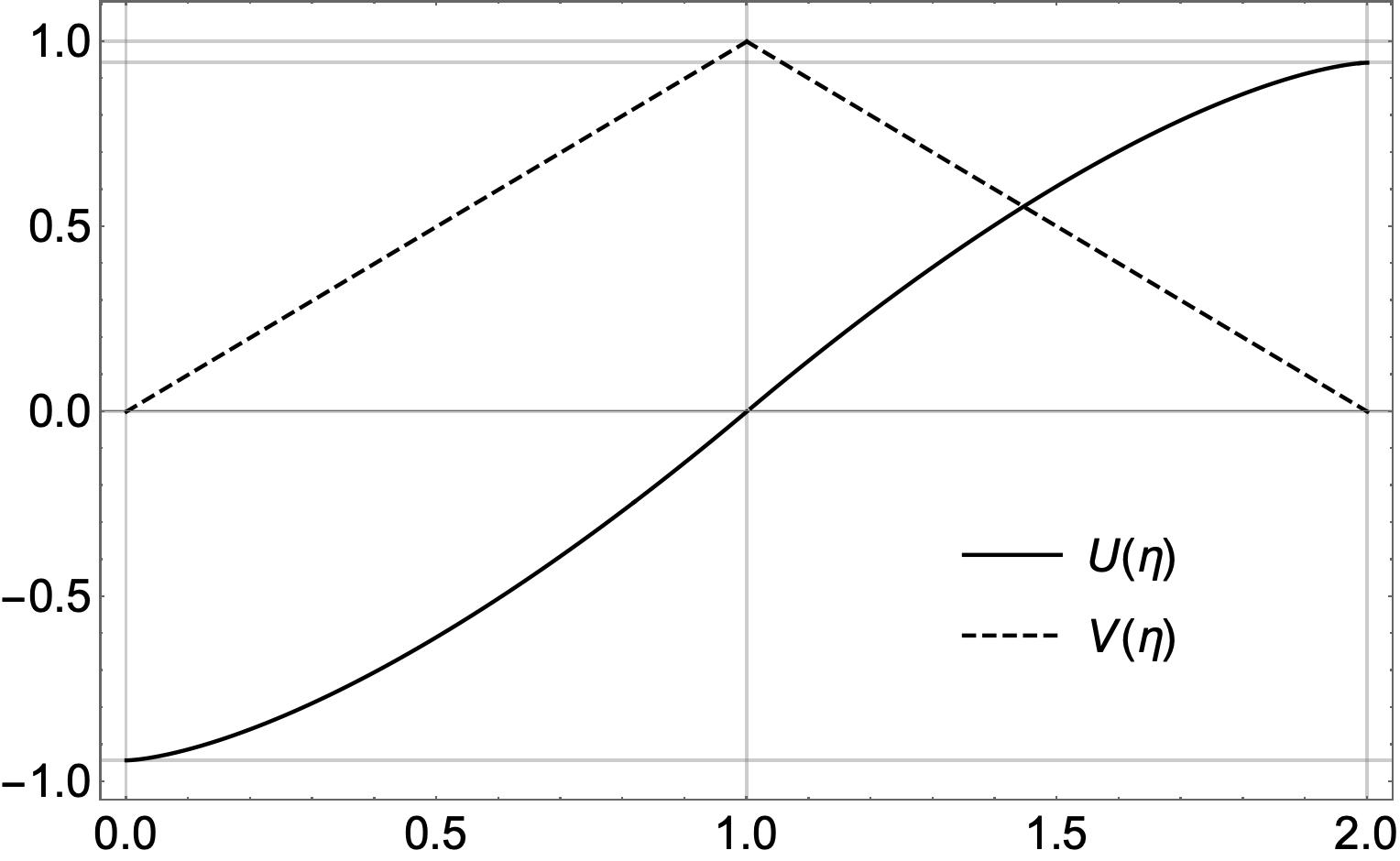}}\hspace{0.5cm}
    \caption{(a) Kink and antikink. (b) The potential $V$ and the pre-potential $U$ on the segment $0<\eta<2$.}
    \label{fig:kak}
\end{figure}

The initial conditions which results in exact oscillons in the signum-Gordon model do not lead to exact oscillons in our periodic model. In order to make oscillons exact and guarantee the existence of kinks we consider a simplified periodic model which shares some characteristics of the signum-Gordon model and the  model with periodic potential. The saw-shape form of the potential
\begin{equation*}
    V(\eta)=\sum_{n=-\infty}^{\infty}|\eta-2n|h_n(\eta)
\end{equation*}
is shown in Fig.~\ref{fig:saw}, where $h_n(\eta):=\theta(\eta-2n+1)-\theta(\eta-2n-1)$.
Its derivative is constant on compact supports. As before, in order to include physically relevant (stable) vacuum solutions $\eta_s=2n$, we assume that $[V'](2n):=0$ and otherwise $[V'](\eta)=\pm 1$. Note that for completeness one can also assume  $[V'](2n+1):=0$ for $\eta_u=2n+1$, however, this static solution is unstable because it corresponds with sharp maxima of the potential. The Euler-Lagrange equation is of the form
\[
    (\partial^2_t-\partial^2_x)\eta+[V'](\eta)=0.
\]
In the vicinity of minima $\eta=2n+\epsilon$, $|\epsilon|<1$, the field $\epsilon$  corresponds with the signum-Gordon field. It means that there exist exact oscillons in our toy model. Of course, there is a limitation for the maximal amplitude of such oscillons.
The presence of degenerate minima  allows for the existence of kinks and antikinks. Such kinks connect two adjacent minima. Here, we look at the case of two minima $\eta=0$ and $\eta=2$. The static kink and antikink solutions are presented in Fig.~\ref{fig:kak}(a). Both solutions are compact and can be translated in space, so we write its solutions in terms of $y=x-a$. The kink $\eta_{K}$ consists on two parabolic pieces that match at $y=\sqrt{2}$
\begin{align*}
    \eta_{K}(y) & =
    \begin{cases}
        0                                     & \text{for } y < 0                          \\
        \frac{1}{2}y^2                        & \text{for } 0 \leq y \leq \sqrt{2}         \\
        2- \frac{1}{2}\Big(y-2\sqrt{2}\Big)^2 & \text{for } \sqrt{2} \leq y \leq 2\sqrt{2} \\
        2                                     & \text{for }  2\sqrt{2}< y
    \end{cases}.
\end{align*}
The antikink is given by $\eta_{\bar K}(y)=2-\eta_{K}(y)$. Both static kink and antikink are BPS solutions that satisfy the first order Bogomolny equations
\[
    \eta'_K(y)=+\sqrt{2V(\eta_K)},\qquad \eta'_{\bar K}(y)=-\sqrt{2V(\eta_{\bar K})}.
\]
The kink and antikink have energy
\begin{equation*}
    E=\pm \int_0^{2\sqrt{2}}dy\; \eta'_{K/\bar K}(y)\sqrt{2V(\eta_{K/\bar K})}=\frac{4\sqrt{2}}{3}
\end{equation*}
which equals to absolute value of the topological charge
\[
    Q:=\int_0^{2\sqrt{2}}\frac{dU}{dy}dy=U\Big(\eta_{K/\bar K}(2\sqrt{2})\Big)-U\Big(\eta_{K/\bar K}(0)\Big)=\pm \frac{4\sqrt{2}}{3}
\]
where the pre-potential $U(\eta)$, obtained from $\sqrt{2V}=\frac{dU}{d\eta}$, have the form
\begin{align*}
    U(\eta) =
    \begin{cases}
        \frac{2\sqrt{2}}{3}\Big(\eta^{3/2}-1\Big)     & \text{for } 0 \leq \eta \leq 1 \\
        \frac{2\sqrt{2}}{3}\Big(1-(2-\eta)^{3/2}\Big) & \text{for } 1 \leq \eta \leq 2 \\
    \end{cases}.
\end{align*}
The pre-potential is plotted in Fig.~\ref{fig:kak}(b).

We want to better understand the behavior of small perturbations on the bulk formed by kinks.
For this reason we now consider the initial configuration that consists on a static kink with an oscillon on top of it, namely
\[
    \eta(0,x)=\eta_K(x)+\chi_\text{osc}(0,x),\qquad \partial_t\eta(0,x)=\partial_t\chi_\text{osc}(0,x).
\]
The field configuration $\{\chi_\text{osc}(0,x),\partial_t\chi_\text{osc}(0,x)\}$ can represent an oscillon at rest or an oscillon in uniform motion. In the absence of the kink, this configuration is evolved by the signum-Gordon equation, and thus it would lead to the exact oscillon for $t>0$. Assuming that the initial configuration $\chi_\text{osc}(0,x)$ is small enough we can solve the field equation exactly for the initial stages of evolution. Assuming a solution of the form
\[
    \eta(t,x)=\eta_K(x)+\chi(t,x)
\]
we get that for $0<\eta<2$ it satisfies the equation
\begin{equation}
    (\partial^2_t-\partial^2_x)\eta-{\rm sgn}(\eta-1)=0. \label{condition}
\end{equation}
We restrict our considerations to $\eta(t,x)$ such that ${\rm sgn}(\eta-1)={\rm sgn}(\eta_K-1)$. It means that
\[
    -\underbrace{\Big(\partial^2_x\eta_K(x)+{\rm sgn}(\eta_K(x)-1)\Big)}_0+(\partial^2_t-\partial^2_x)\chi(t,x)=0
\]
i.e.\ $\chi(t,x)$ obeys the wave equation and has the form $\chi(t,x)=F(x+t)+G(x-t)$.
The condition \eqref{condition} is valid only for certain $t<t_\text{max}$. Moreover, since the function $\eta_K(x)-1$ changes sign at $x=\sqrt{2}$ (center of the kink), we must centralize the oscillon at  $x=\frac{\sqrt{2}}{2}$ or $x=\frac{3\sqrt{2}}{2}$.  Denoting for simplicity  $f(x)\equiv \chi_\text{osc}(0,x)$ and $g(x)\equiv \partial_t\chi_\text{osc}(0,x)$ we get the solution of the Cauchy problem in the form given by d'Alembert formula
\begin{equation}
    \chi(t,x)=\frac{1}{2}\Big(f(x+t)+f(x-t)\Big)+\frac{1}{2}\int_{x-t}^{x+t}ds\,g(s).\label{solchi}
\end{equation}
By assumption the functions $f(x)$ and $g(x)$ have compact supports.

First we will look at the case of an oscillon at rest. If the initial configuration coincides with the oscillon having phase such that $g(x)=0$ and $f(x)$ has maximum amplitude, the function $f(x)$ is given by three partial function $f_1(x), f_2(x), f_3(x)$ such that
\[
f_k(x)=\epsilon^2\phi_k\left(\frac{x-\frac{\sqrt{2}}{2}}{\epsilon}+\frac{1}{2}\right)
\]
where $\phi_k$ are partial expressions of the oscillon profile of unit size.
The center of a rescaled  oscillon is localized at $x=\frac{\sqrt{2}}{2}$ which lies at half the distance from the kink left border to its center. The scaling transformation allows to obtain an oscillon with support localized totally inside kink support.
For small values of $\epsilon$ the initial profile and the solution satisfy the condition \eqref{condition}. The functions $\phi_k(x)$, $k=1,2,3$ are given by expressions
\[\phi_1(x)=-\frac{x^2}{2}, \qquad \phi_2(x)=\frac{x^2}{2}-\frac{x}{2}+\frac{1}{16},\qquad \phi_3(x)=-\frac{1}{2}(1-x)^2.\]
The function $\phi_1(x)$ has support $x\in(0,\frac{1}{4})$, $\phi_2(x)$ has support $x\in(\frac{1}{4},\frac{3}{4})$ and $\phi_3(x)$ is given on $x\in(\frac{3}{4},1)$. Such initial condition  represents the oscillon shape $\phi(t,x)$ at $t=\frac{1}{4}$, i.e.\ at the instant of time when time derivative $\partial_t\phi(t,x)|_{t=1/4}$ vanishes everywhere, see \cite{Arodz:2007jh}.
The solution consists on a kink and two bumps $\frac{1}{2}f(x\pm t)$ that moves with the speed of light in opposite directions. This is true as long as the condition \eqref{condition} holds. One can conclude that the initial condition which would lead to the oscillon actually does not lead  to it because the perturbation is governed by the wave equation. The very characteristic property of this solution is splitting the perturbation into two shapes that moves in opposite directions.

Now we will look at a perturbation of more complicated form. Namely, we choose an initial condition which would lead to an oscillon moving with velocity $V$ in the positive direction of the $x$ axis, provided that the field equation is the signum-Gordon equation. The oscillon in motion can be obtained by applying a Lorentz boost. Such oscillon is shorter than the unit oscillon at rest, i.e.\ it has size $\gamma^{-1}=\sqrt{1-V^2}$. The number and type of partial solutions that form the oscillon depends on the value of velocity $V$. The analysis of all possibilities is out of scope of this paper. We are interested in one example of analytical solution that can be compared with the numerical simulations. For this reason we choose $V=\frac{3}{4}$ which gives $\gamma^{-1}=\frac{\sqrt{7}}{4}$. The Lorentz boosted oscillon at $t=0$ consists on five nontrivial partial solutions matched at
\begin{equation}
{x_1=\frac{2}{7}\gamma^{-1},\quad x_2=\frac{4}{7}\gamma^{-1},\quad x_3=\frac{2}{3}\gamma^{-1},\quad x_4=\frac{6}{7}\gamma^{-1}.\label{xk}}
\end{equation} At $x_0=0$ and $x_5=\gamma^{-1}$
the oscillon matches the vacuum solution of the signum-Gordon field $\varphi=0$. The field at $t=0$, $\varphi_k(0,x)$ and its derivative $\partial_t\varphi_k(0,x)$ have the form
\begin{align*}
    & \varphi_1(0,x)=\frac{15}{14}x^2                                   &  & \partial_t \varphi_1(0,x)=-\frac{13}{7}x                             &  & x_0<x<x_1\nonumber \\
    & \varphi_2(0,x)=-\frac{17}{7}x^2+\frac{\sqrt{7}}{2}x-\frac{1}{8}   &  & \partial_t \varphi_2(0,x)=\frac{36}{7}x-\frac{\sqrt{7}}{2}           &  & x_1<x<x_2\nonumber \\
    & \varphi_3(0,x)=\frac{15}{14}x^2-\frac{\sqrt{7}}{2}x+\frac{3}{8}   &  & \partial_t \varphi_3(0,x)=-\frac{13}{7}x+\frac{\sqrt{7}}{2}          &  & x_2<x<x_3\nonumber \\
    & \varphi_4(0,x)=\frac{33}{14}x^2-\frac{13}{2\sqrt{7}}x+\frac{5}{8} &  & \partial_t \varphi_4(0,x)=-\frac{37}{7}x+\frac{15}{2\sqrt{7}}        &  & x_3<x<x_4\nonumber \\
    & \varphi_5(0,x)=-\frac{8}{7}\Big(x-\frac{\sqrt{7}}{4}\Big)^2       &  & \partial_t \varphi_5(0,x)=\frac{12}{7}\Big(x-\frac{\sqrt{7}}{4}\Big) &  & x_4<x<x_5\nonumber
\end{align*}
The form of perturbation $f(x)$, $g(x)$ is obtained as a combination of a scaling transformation $\varphi_{\epsilon}(t,x)=\epsilon^2\varphi(\frac{t}{\epsilon},\frac{x}{\epsilon})$ and translation of the Lorentz boosted solution by $x_c=\frac{\sqrt{2}}{2}-\frac{\epsilon}{2\gamma}$. This translation put the center of the oscillon at half distance from the left border of the kink to its center. The functions representing initial profile of the perturbation and its time derivative at $t=0$ have the form
\begin{align}
    f_k(x) & =\epsilon^2\;\varphi_{\epsilon,k}(0,x-x_c)=\epsilon^2\;\varphi_k\Big(0,\frac{x-\frac{\sqrt{2}}{2}}{\epsilon}+\frac{1}{2\gamma}\Big),    \label{f}     \\
    g_k(x) & =\epsilon\;\partial_t\varphi_{\epsilon,k}(0,x-x_c)=\epsilon\;\partial_t\varphi_k\Big(0,\frac{x-\frac{\sqrt{2}}{2}}{\epsilon}+\frac{1}{2\gamma}\Big).\label{g}
\end{align}
where $k=1,2,\ldots 5$ and $\frac{1}{2\gamma}=\frac{\sqrt{7}}{8}$.  The matching points read
\[
\widetilde x_k=x_c+\epsilon\, x_k=\frac{\sqrt{2}}{2}-\frac{\epsilon}{2\gamma}+\epsilon\, x_k,\qquad k=0,1,\ldots,5\]
where $x_k$ are given by \eqref{xk}.

\begin{figure}[ht!]
    \subfigure
    []
    {\includegraphics[width=0.45\textwidth,height=0.25\textwidth]{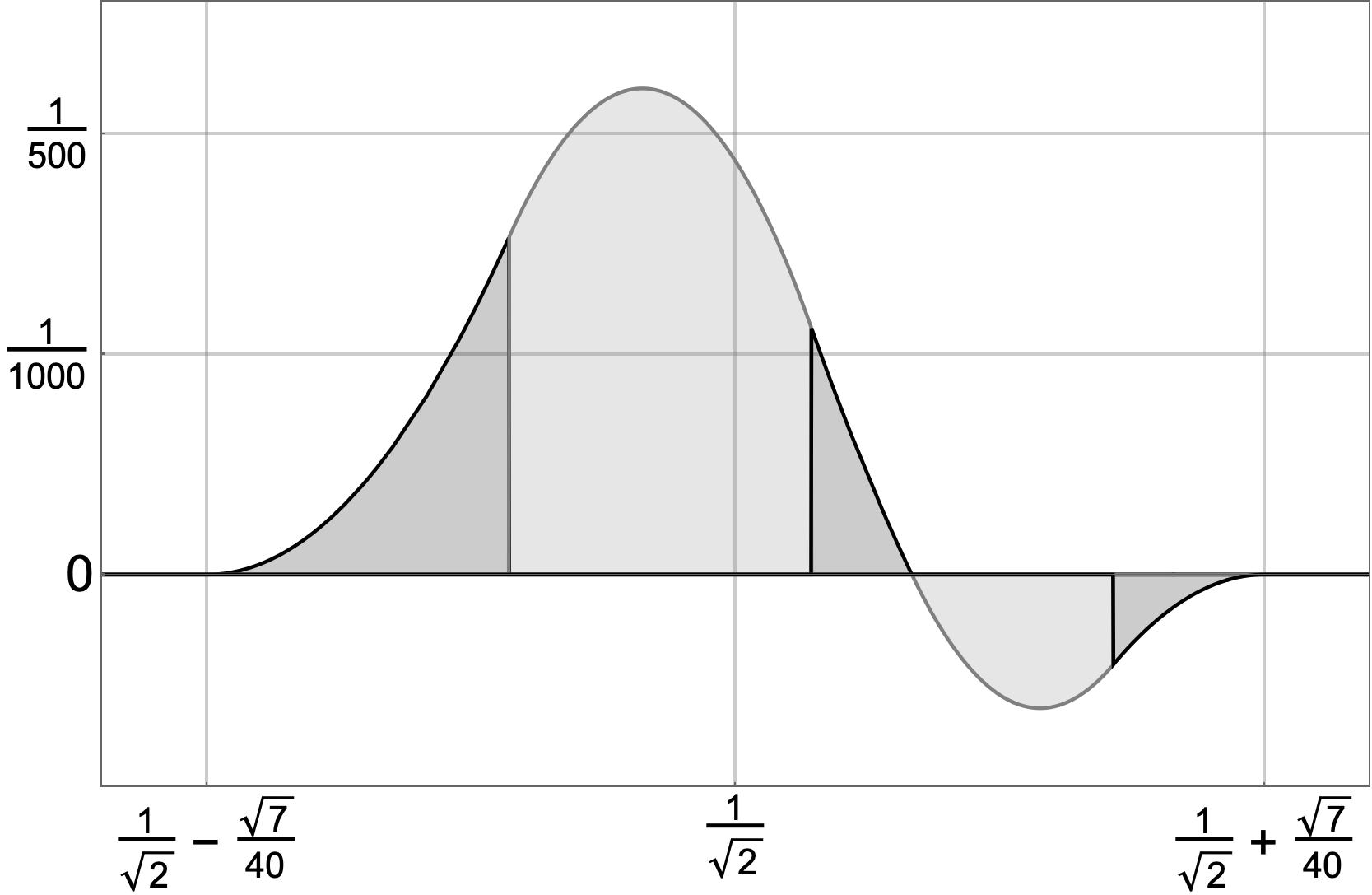}}\hspace{0.5cm}
    \subfigure
    []
    {\includegraphics[width=0.45\textwidth,height=0.25\textwidth]{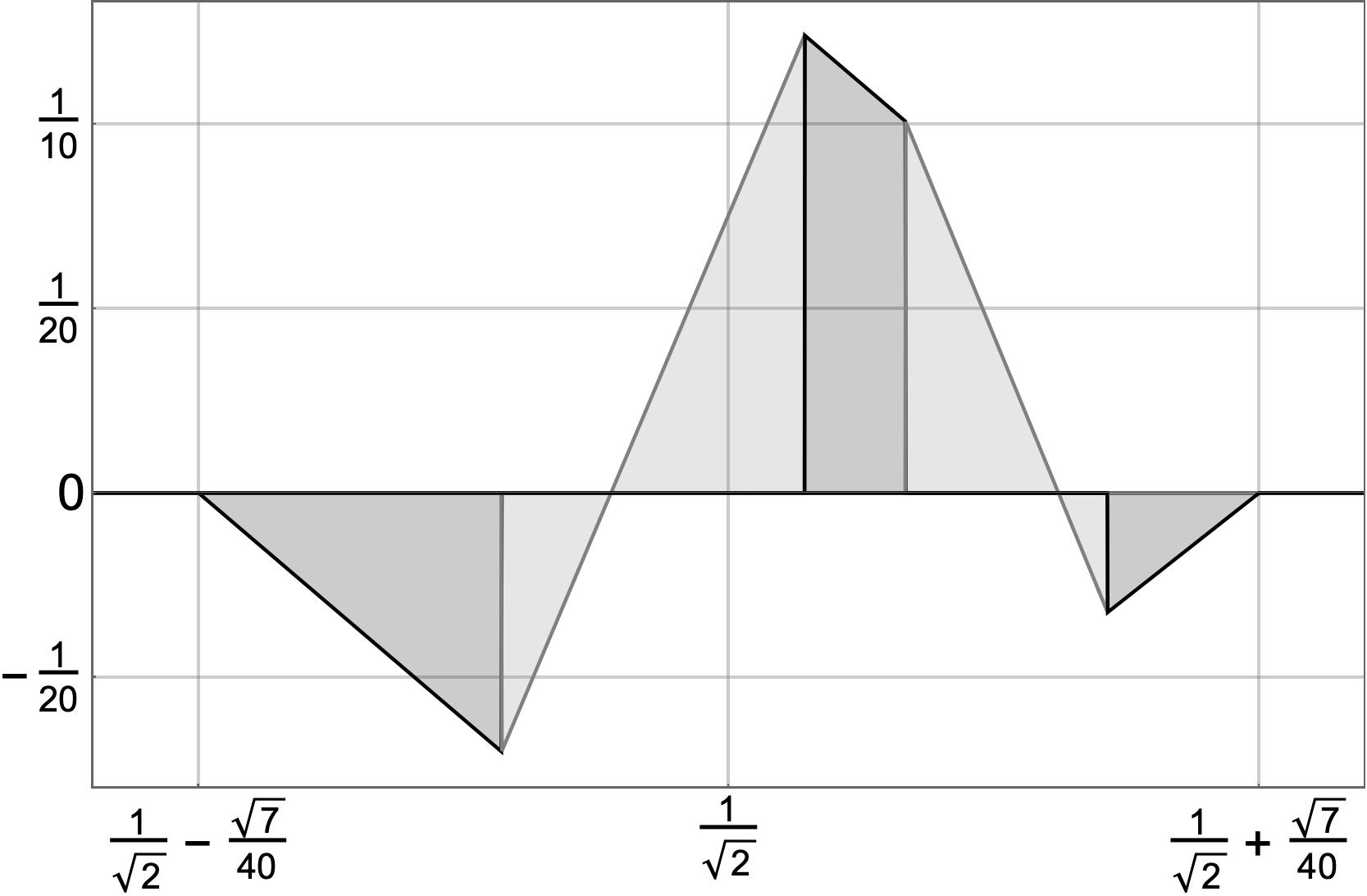}}\hspace{0.5cm}
    \caption{The oscillon at $t=0$ centralized at $x=\frac{\sqrt{2}}{2}$ for  $\epsilon=\frac{1}{5}$. (a) the oscillon shape, (b) time derivative of the initial profile.}
    \label{fig:oscini}
\end{figure}

\begin{figure}[ht!]
    \subfigure
    []
    {\includegraphics[width=0.45\textwidth,height=0.25\textwidth]{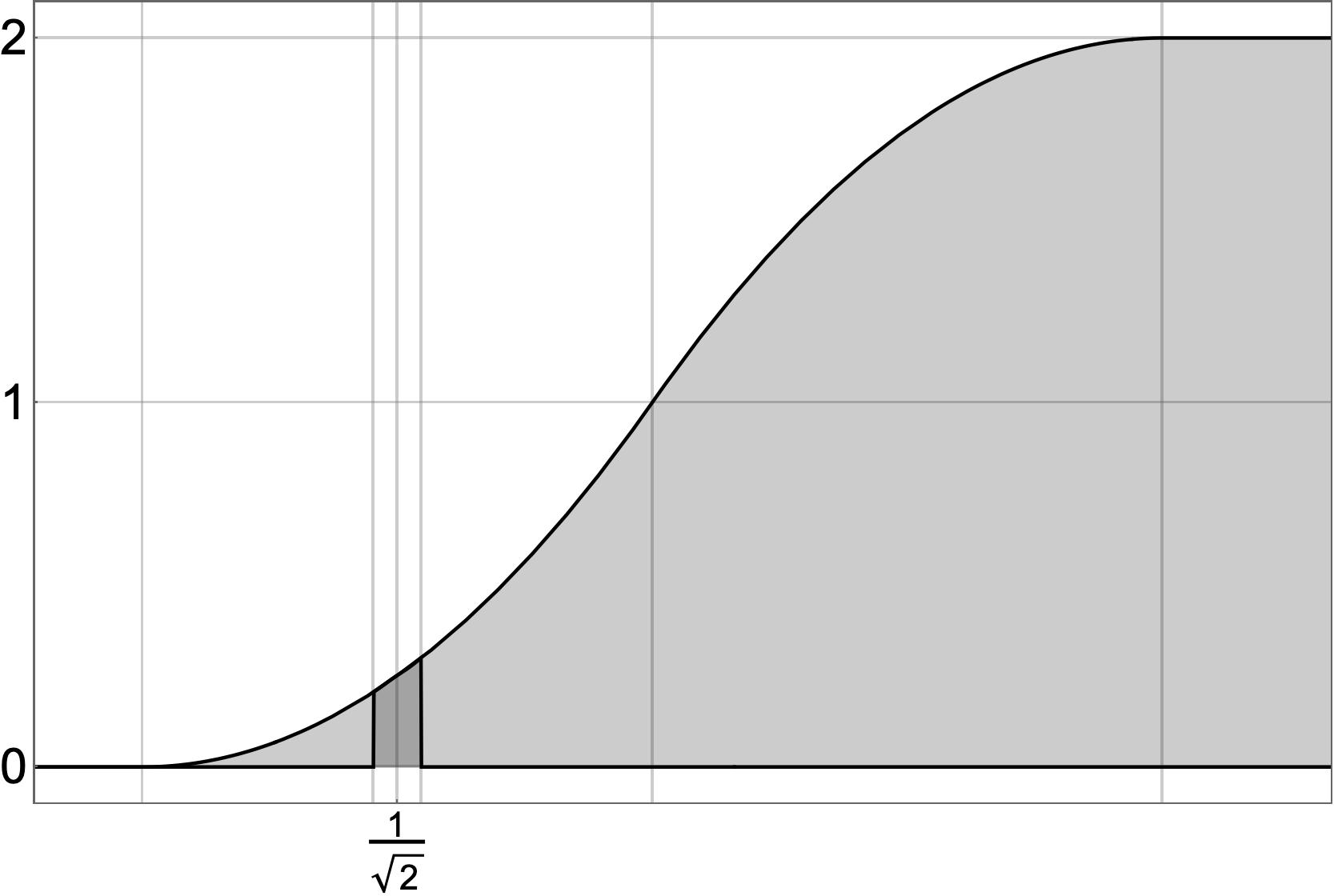}}\hspace{0.5cm}
    \subfigure
    []
    {\includegraphics[width=0.45\textwidth,height=0.25\textwidth]{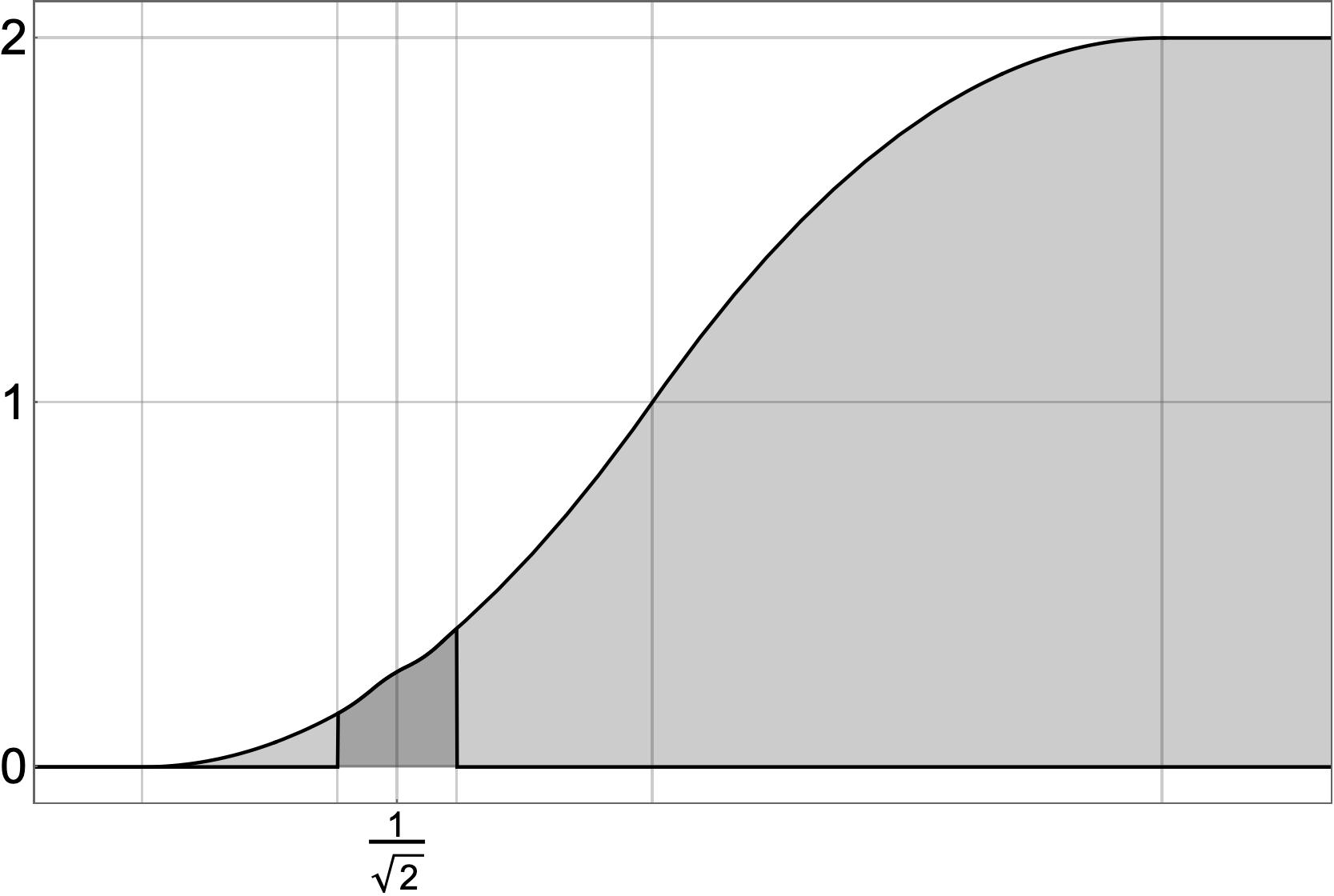}}\hspace{0.5cm}
    \caption{The kink with the oscillon on top of it at $t=0$. (a) $\epsilon=\frac{1}{5}$, (b) $\epsilon=\frac{1}{2}$}
    \label{fig:kinkplusosc}
\end{figure}

\begin{figure}[ht!]
    {\includegraphics[width=0.55\textwidth,height=0.35\textwidth]{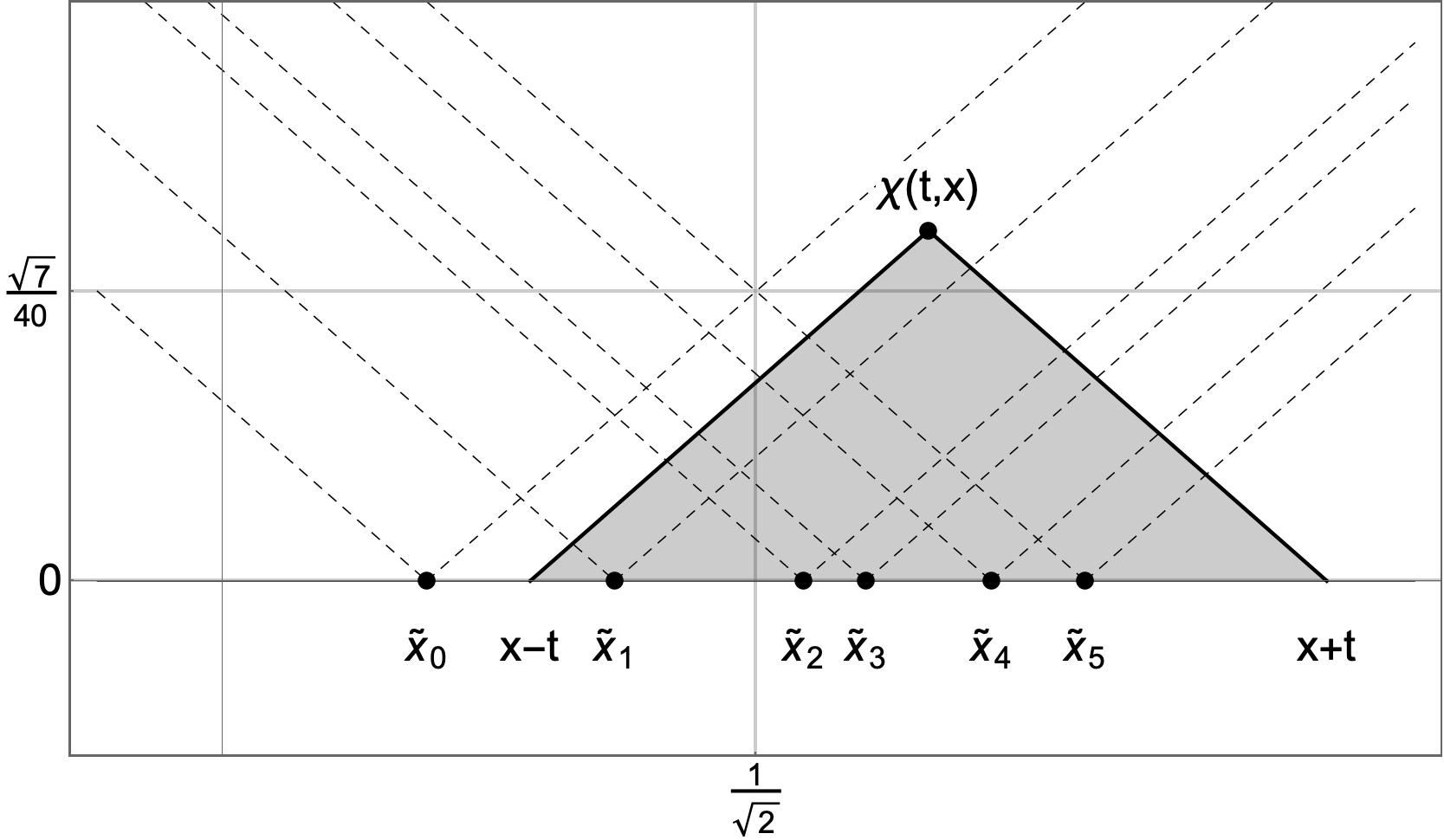}}\hspace{0.5cm}
    \caption{The causal structure of the solution obtained from the initial data on the segment $[\widetilde x_0,\widetilde x_5]$. In the region of intersection of two stripes $x-\widetilde x_5<t<x-\widetilde x_0$ and $-x+\widetilde x_0<t<-x+\widetilde x_5$ the solution is a superposition of functions that depend on $x-t$ and $x+t$. In the right stripe above $t=-x+\widetilde x_5$ the solution is exclusively a function of $x-t$. Similarly, in the left stripe above $t=x-\widetilde x_0$ the solution is a function of $x+t$. In the region above the stripes the solution is constant and below the strips the solution correspond with the vacuum solution.}
    \label{fig:cones}
\end{figure}

In Fig.~\ref{fig:oscini} we show the initial profile of the perturbation (a) and its time derivative (b) whereas in Fig.~\ref{fig:kinkplusosc} we present the initial field configuration that consists of kink and perturbation.  The perturbation is evolved by the wave equation. For this reason the solution $\chi(t,x)$ does not behave as the oscillon in the signum-Gordon model. The solution is actually given by \eqref{solchi}. The causality in $1+1$ dimension implies that the solution $\chi(t,x)$ depends on the initial data localized inside the past light-cone of the event $(t,x)$.  Fig.~\ref{fig:cones} shows the spacetime diagram representing the propagation of the initial data. The light-cone lines were plotted for each point $\widetilde x_k$ at which different partial solutions match. The solution can be explicitly determined for any event $(t,x)$ using d'Alembert formula. Here we shall present the solution for $t>\frac{\epsilon}{2\gamma}$, i.e.\ for times when the solution consists on two non-overlapping functions of $x-t$ and $x+t$.

The solution $\chi(t,x)$ has different form in different regions. Let us consider a horizontal line with $t>\frac{\epsilon}{2\gamma}$. The solution consists on partial solutions belonging to the left stripe $\chi_{L_k}(t,x)$, $k=1,\ldots,5$ (functions of $x+t$),  the central partial solution  $\chi_{C}(t,x)$, and partial solutions belonging to the right stripe $\chi_{R_k}(t,x)$, $k=1,\ldots,5$ (functions of $x-t$).
The central solution is constant because the function $f(s)$, with $s=x+t$ and $s=x-t$,  is taken for $s$ outside the support of the initial data and thus $f(s)=0$. Also, the integration of $g(s)$ is performed over a complete compacton support giving
\begin{equation*}
    \chi_C(t,x) = \frac{1}{2}\sum_{k=1}^5\int_{\widetilde x_{k-1}}^{\widetilde x_{k}}g_k(s)ds=\frac{\epsilon^2}{48}
\end{equation*}
where $f_k(s)$ and $g_k(s)$ are given by \eqref{f} and \eqref{g}.
For the solutions inside the right stripe one gets
\begin{align*}
    \chi_{R_j}(t,x)&=\frac{1}{2}f_j(x-t)+\frac{1}{2}\int_{{x-t}}^{\widetilde x_{j}}g_j(s)ds+\frac{1}{2}\sum_{k> j}\int_{\widetilde x_{k-1}}^{\widetilde x_{k}}g_k(s)ds\\
    &=A_{R_j}(x-t)^2+B_{R_j}(x-t)+C_{R_j}, \qquad j,k=1,\ldots,5.
\end{align*}
Similarly, the partial solutions that form the left stripe are obtained as
\begin{align*}
\chi_{L_j}(t,x)&=\frac{1}{2}f_{6-j}(x+t)+\frac{1}{2}\sum_{k\le5-j}\int_{\widetilde x_{k-1}}^{\widetilde x_k}g_k(s)ds+\frac{1}{2}\int_{\widetilde x_{5-j}}^{ x+t}g_{6-j}(s)ds\\
&=A_{L_j}(x+t)^2+B_{L_j}(x+t)+C_{L_j} , \qquad j,k=1,\ldots,5.
\end{align*}
All partial solutions are quadratic functions of $x-t$ and $x+t$. Their coefficients are presented in appendix~\ref{app:partial-solutions}.
\begin{figure}[ht!]
    \subfigure
    []
    {\includegraphics[width=0.45\textwidth,height=0.25\textwidth]{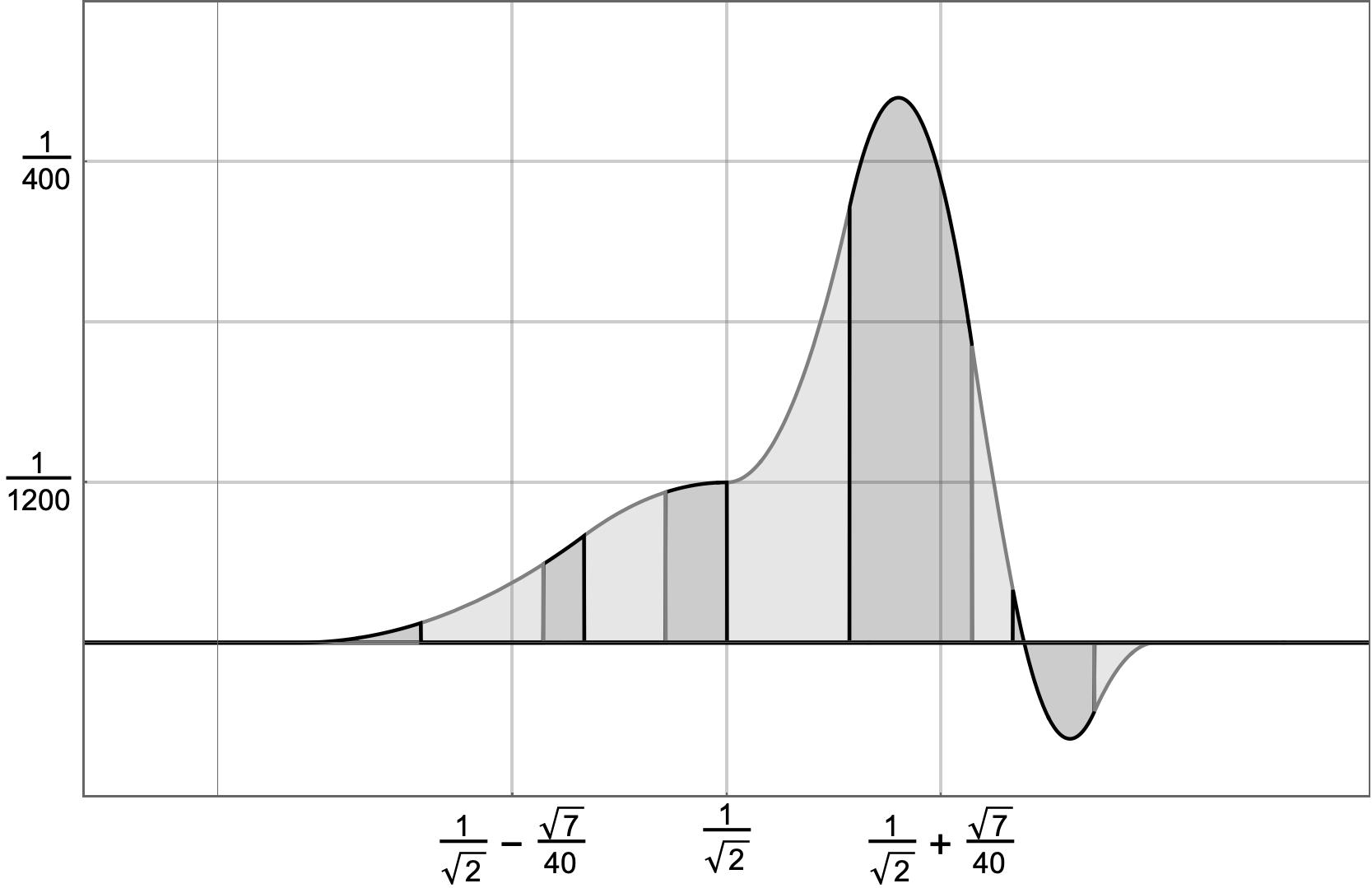}}\hspace{0.5cm}
    \subfigure
    []
    {\includegraphics[width=0.45\textwidth,height=0.25\textwidth]{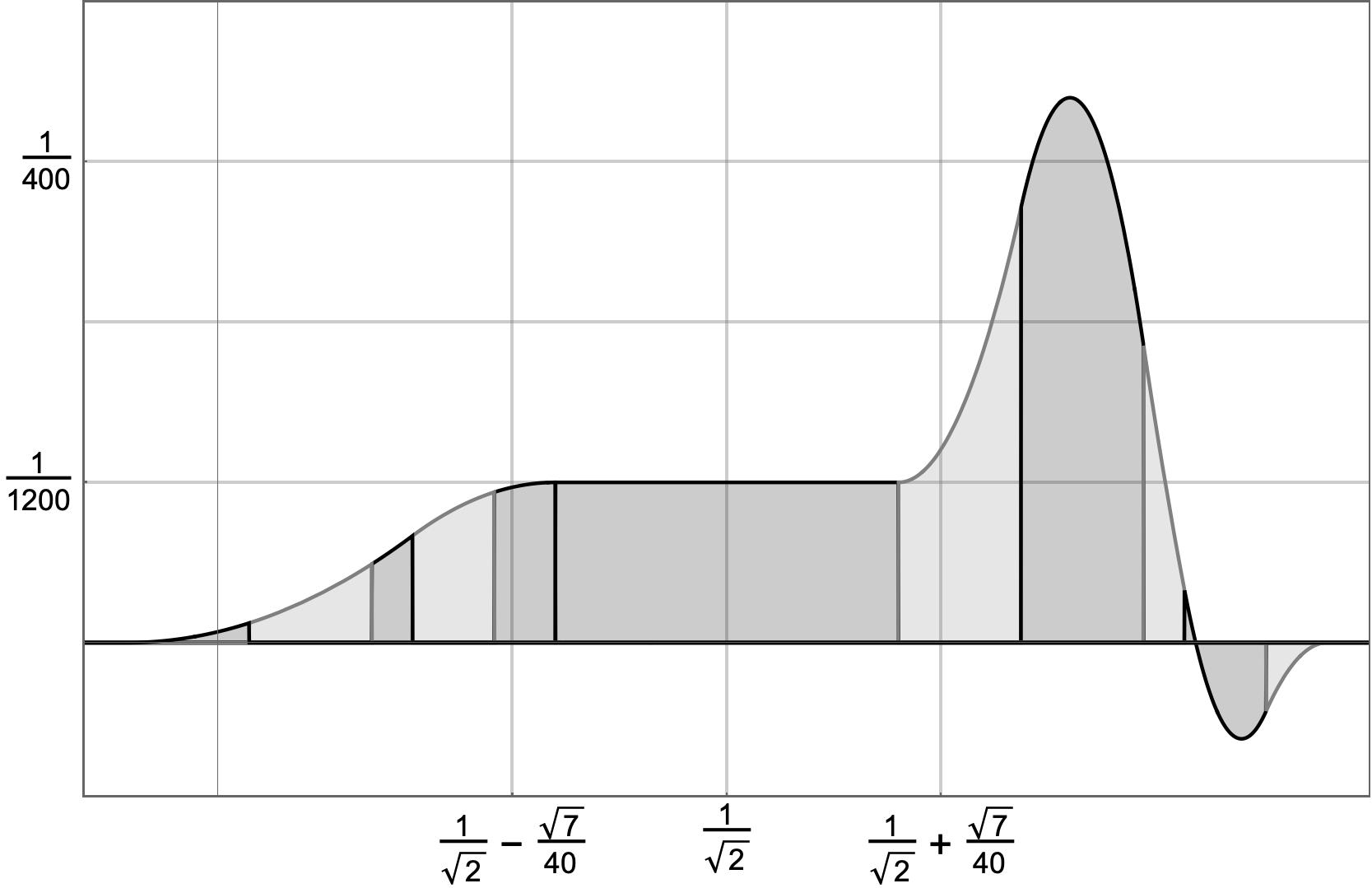}}\hspace{0.5cm}
    \caption{Evolution of the initial profile with $\epsilon=\frac{1}{5}$ at  (a) $t=\frac{\epsilon}{2\gamma}$, (b) $1.8 \,\frac{\epsilon}{2\gamma}$. The initial data have support $x\in[\frac{1}{\sqrt{2}}-\frac{7}{40},\frac{1}{\sqrt{2}}+\frac{7}{40}]$. The perturbation at $t=1.8\,\,\frac{\epsilon}{2\gamma}$ has support $x\in[\frac{1}{\sqrt{2}}-\frac{(1+1.8)\epsilon}{2\gamma},\frac{1}{\sqrt{2}}+\frac{(1+1.8)\epsilon}{2\gamma}]\approx[0.52,0.89]$ which is localized inside the kink support.}
    \label{fig:evolution}
\end{figure}

An important point about the present analysis is that the solution describing propagation of the oscillon profile  can be obtained exactly (in its domain of validity). It allows to compare the analytical results with numerical simulation.
Note that  the initial oscillon would propagate with velocity $V$ in the signum-Gordon model, however, on the bulk of static kink  it propagates  in opposite directions with the speed of light. We can expect that oscillons which collide with kinks would be transferred to the other side of kinks with higher speed that the initial speed of the oscillon before collision. Of course,  in the initial stage of collision the oscillon can be perturbed and thus the profile that would propagate on the bulk of kink would not correspond with initial data for the exact oscillon.

\begin{figure}
    \includegraphics{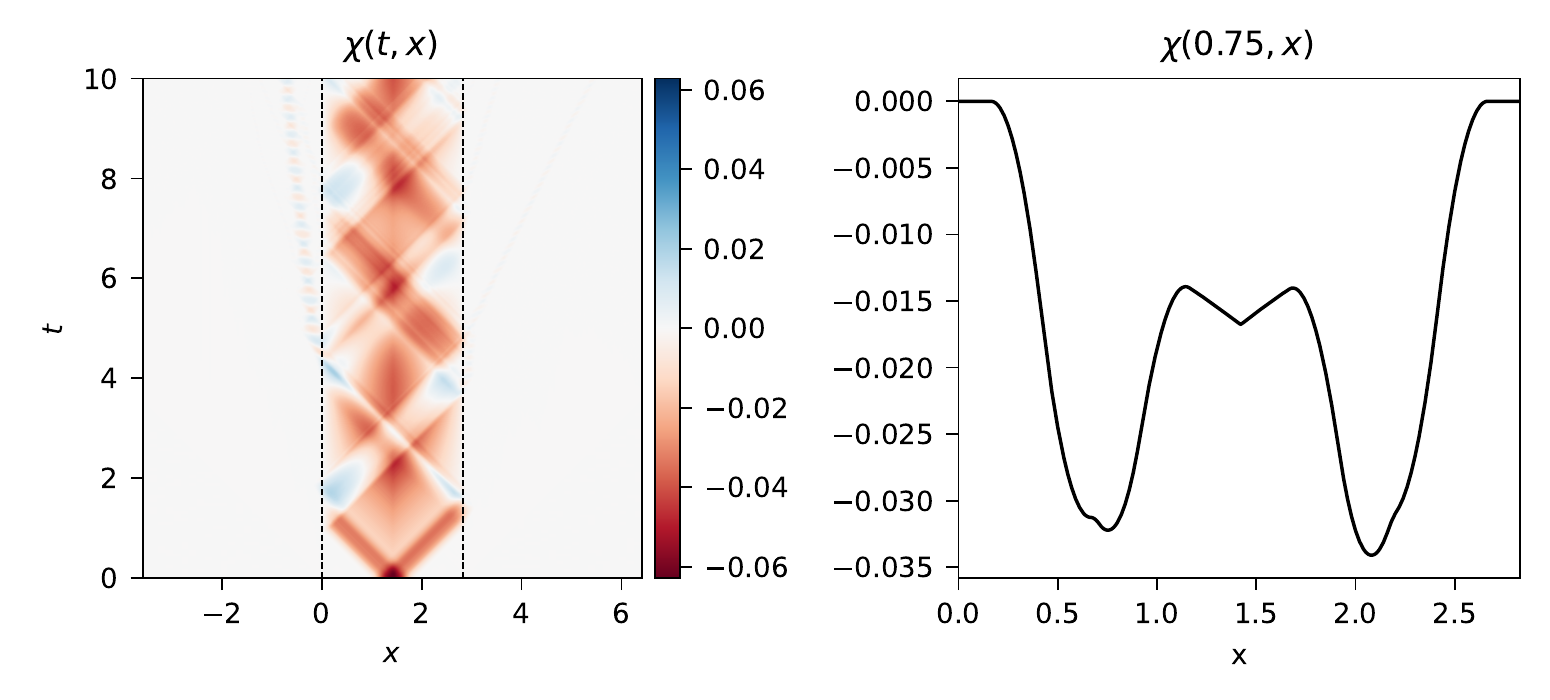}
    \caption{Perturbation $\chi(t, x)$ (left) for the case of oscillon with parameters $l = 1$, $V = 0$, $\alpha = 0.25$, $v_0 = 0$ added on top of a kink. Oscillon and kink have aligned central points. The oscillon placement creates an asymmetry in the perturbation profile (left) before it can reach the kink borders.}
    \label{fig:toy/centered}
\end{figure}

We will now present some numerical results for the toy-model.
Once again we considered an oscillon profile added on top of the kink.
Figure~\ref{fig:toy/centered} shows the results for the case where the oscillon has its middle point aligned with the kink center.
Since $\eta_K(\sqrt{2}) = 1$ the condition $\sgn(\eta-1)=\sgn(\eta_K-1)$ is not satisfied around this point and an otherwise symmetric perturbation $\chi(t,x)$ becomes asymmetric before reaching the kink borders.
This behavior is unique to the toy-model, and it is a consequence of the sharp maximum of the potential.

\begin{figure}
    \includegraphics{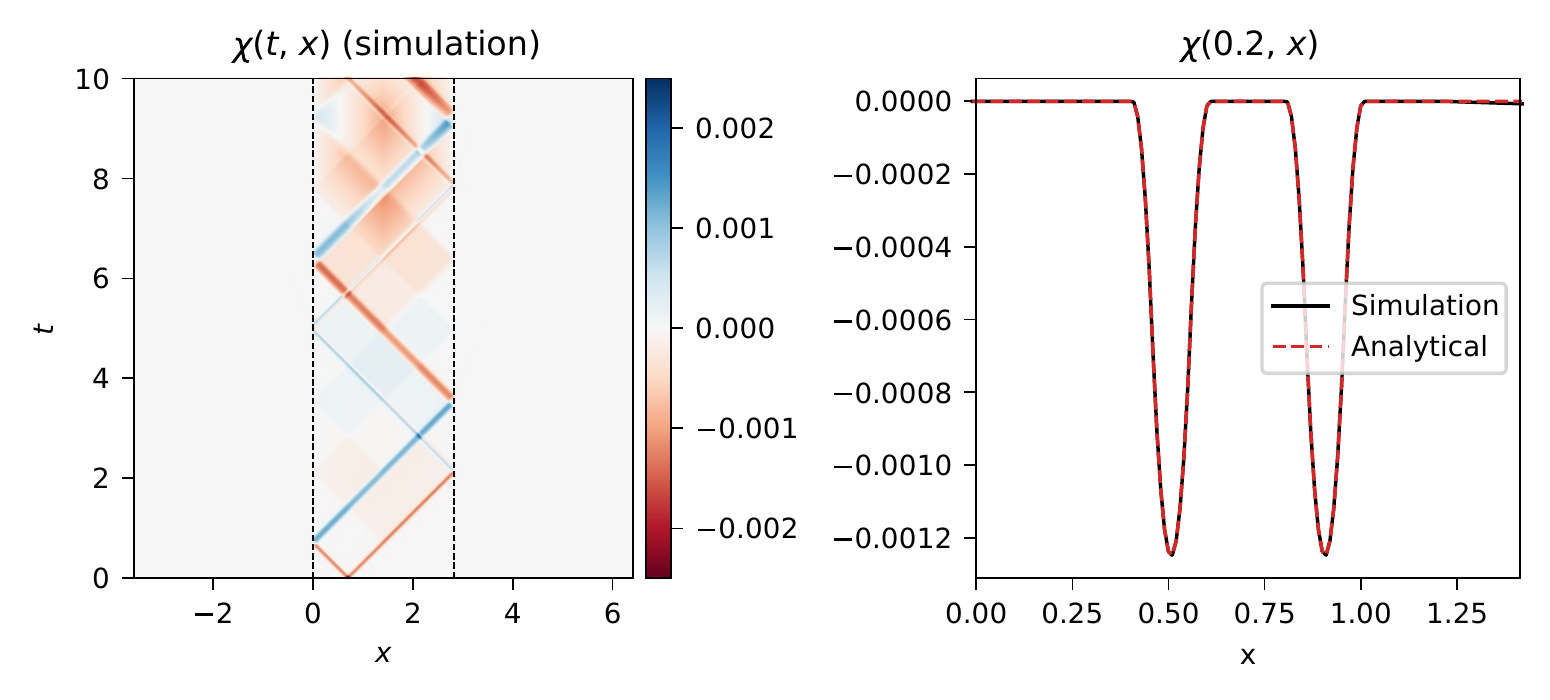}
    \caption{Evolution of initial configuration formed by  ``oscillon at rest''. Comparison between simulation and analytical result for $l = 1$, $V = 0$, $\alpha = 0.25$, and $v_0 = 0$. Oscillon initially centered around $x = \sqrt{2}/2$.}
    \label{fig:toy/quarter-rest}
\end{figure}

\begin{figure}
    \includegraphics{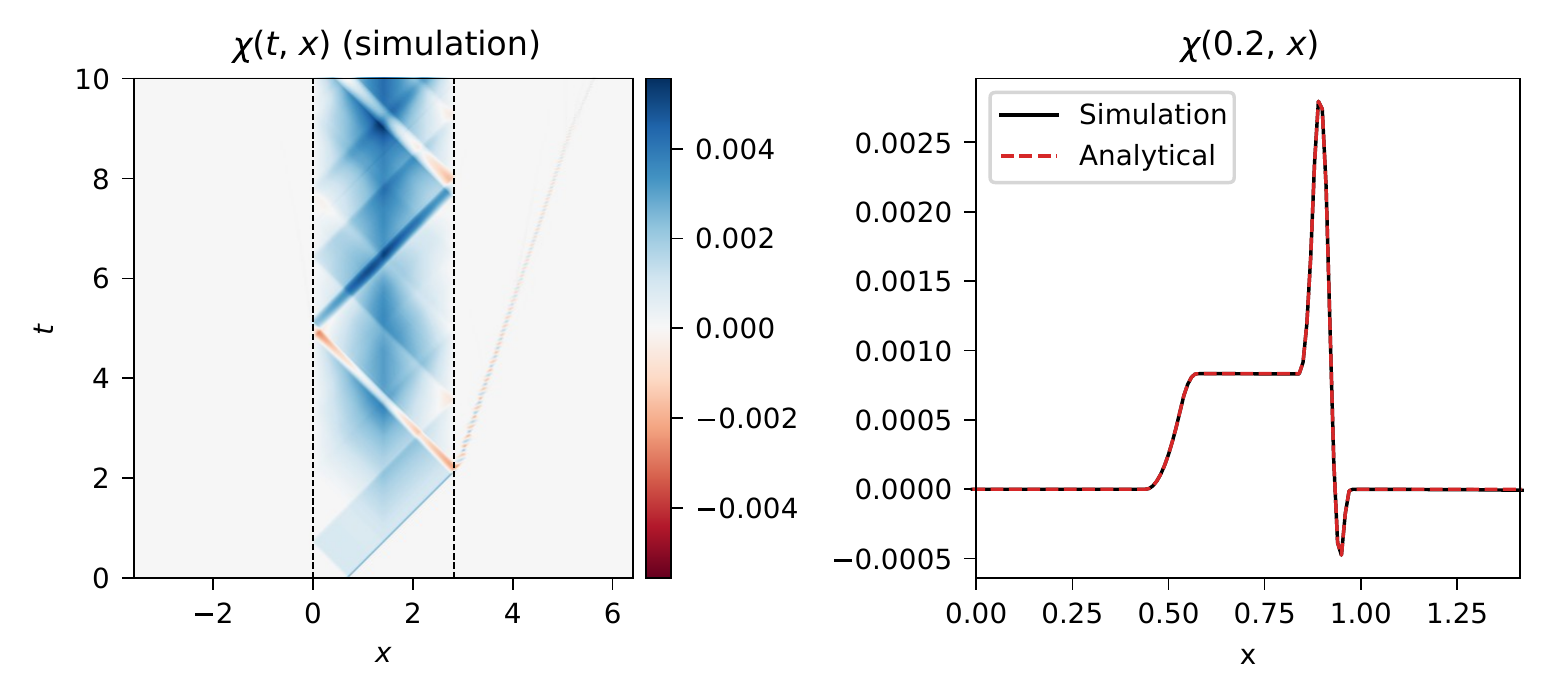}
    \caption{Evolution of initial configuration formed by ``oscillon in motion''. Comparison between simulation and analytical result for $l = 1$, $V = 0.75$, $\alpha = 0$, and $v_0 = 0$. Oscillon initially centered around $x = \sqrt{2}/2$.}
    \label{fig:toy/quarter-moving}
\end{figure}

To satisfy the condition $\sgn(\eta-1)=\sgn(\eta_K-1)$, we consider cases where the oscillon has its middle point aligned with $x = \frac{\sqrt{2}}{2}$, halfway between the kink left border and its center.
We present simulation results for $V=0$ and $V=0.75$ and compare them with the analytical expressions in Fig.~\ref{fig:toy/quarter-rest}--\ref{fig:toy/quarter-moving}.
Both approaches yield equal results in its domains of validity.

\section{Generalized model}
\label{sec:generalized}

In the previous section we claimed that the sharpness at potential maxima causes an asymmetry in an otherwise symmetric perturbation.
Let us expand on this by considering a generalized model with potential
\begin{equation*}
    V_k(\eta) = \sum_{n=-\infty}^\infty \frac{2-k}{2} \left( 1 - |x - 2n - 1|^{1+k} \right) h_n(\eta).
\end{equation*}
This potential is such that for $k=0$ it reproduces the toy-model and for $k=1$ the parabolic potential initially considered.
For values of $k$ different from odd integers, the potential is sharp at its maxima $\eta = 2n + 1$, $n \in \mathbb{Z}$.
The generalized model has minima for $\eta$ equals to even integers, and static kink solutions obeying BPS equation.
In general, the BPS is separable, and the kink solutions $\eta(x)$ can be obtained from the transcendental equation
\begin{align}
    \frac{-1 + \eta_K(x)}{\sqrt{2 - k}} {}_2F_1 \left(\frac{1}{2}, \frac{1}{1 + k}, 1 + \frac{1}{1 + k}; |1 - \eta_K(x)|^{1 + k} \right) = x - x_0(k) \label{eq:kink-transcendental}
\end{align}
where ${}_2F_1(a, b, c; z)$ is the hypergeometric function and
\begin{equation*}
    x_0(k) = \sqrt{\frac{\pi}{2 - k}} \frac{\Gamma\left(1 + \frac{1}{1 + k}\right)}{\Gamma\left(\frac{1}{2} + \frac{1}{1 + k}\right)}
\end{equation*}
is the position of the kink middle point.
The kink support is defined by $x \in [0, 2x_0(k)]$.
In general, the kink profile must be obtained finding the roots of equation~\eqref{eq:kink-transcendental}.

\begin{figure}
    \includegraphics{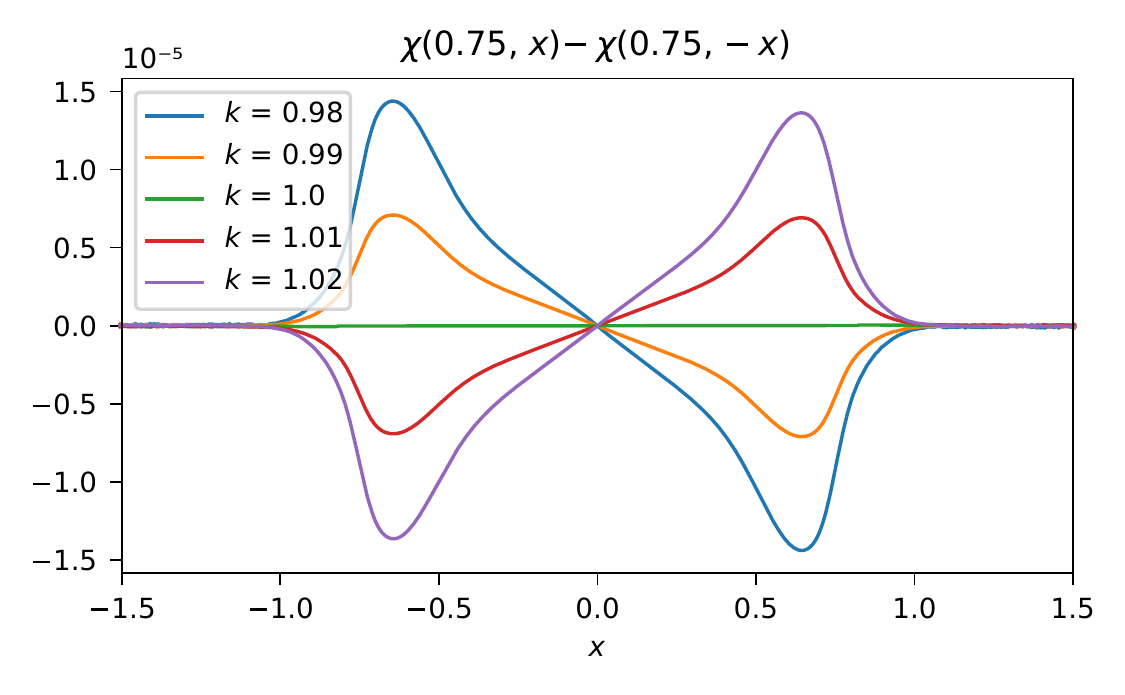}
    \caption{Perturbation asymmetry $\chi(t, x) - \chi(t, -x)$ for kink and oscillon with centers aligned with the point $x=0$. Oscillon parameters are $l = 1$, $V = 0$, $\alpha = 0,25$, $v_0 = 0$.}
    \label{fig:asymmetry}
\end{figure}

To illustrate the consequences of the non-analytical maximum of $V_k(\eta)$, we performed simulations where an oscillon profile was added on top of a kink, with its middle point aligned with the kink center.
The entire configuration was translated so that the kink support was $x \in [-x_0(k), x_0(k)]$.
We considered values of $k$ close to 1, meaning potential similar to the quadratic potential.
For this set of potentials, only $V_0(\eta)$ has analytical maxima.
Figure~\ref{fig:asymmetry} shows that in fact, only the case $k=0$ preserved the perturbation symmetry during the initial stages of propagation.

\section{Conclusions}
\label{sec:conclusions}

In this paper we studied compact kinks in a model with non-analytic potential.
The dynamics of perturbed kinks obtained shrinking and stretching BPS kinks was studied through numerical simulations and analytical arguments.
We were able to effectively describe the vibration of the kink support in terms of a scale transformation together with a perturbation of the kink profile inside its support.
The missing ingredient in the effective description is radiation, consisting primarily of compact oscillons.
We observed that the emission of compact oscillons is an important mechanism to the release of additional energy.
This means that is important to understand the interaction between compact oscillons and compact kinks.

We considered this interaction in the form of a scattering process, where a moving compact oscillon collides with a compact kink.
Our simulations have shown that the oscillon transverses the kink bulk without altering its support in the form of light-speed pulses.
This pulses give origin to new oscillons outside the kink support.
We have also measured the energetic efficiency of this process as a function of the scattering parameters.
In particular, we found a curious non-monotonic dependence on the velocity and a strong dependence on the oscillon phase.

Since much of the scattering dynamics depends on the propagation of signals through the kink bulk, we have also studied the dynamics of localized perturbations added to a kink profile.
We have shown that such perturbations are mostly governed by a Klein-Gordon equation with mass parameter $m^2=-1$.
This observation explains why signals travel with the speed of light through the kink, and also have allowed us to obtain the evolution of the perturbation through a semi-analytical approach.
This new approach yielded consistent results with the simulations.

We were able to obtain closed formulas for the perturbation profile in a toy-model. These expressions were also consistent with numerical results in its domain of validity.
The toy-model also presented an interesting behavior where some otherwise symmetrical perturbations evolved to asymmetric profiles.
This behavior was understood as a consequence of the sharp maxima of the potential.

\begin{acknowledgments}
The authors are indebted to H.~Arod\'z, A.~Wereszczy\'nski and P.~E.~Assis for discussion and valuable remarks.
F.~M.~Hahne is supported by CNPq---Brazil.
\end{acknowledgments}

\appendix
\section{Euler-Lagrange equations for \texorpdfstring{$b(t)$}{b(t)} and \texorpdfstring{$A(t)$}{A(t)}}
\label{app:ELbA}

\begin{multline*}
    \left[\left(\frac{275 \pi }{96}-\frac{\pi ^3}{2}\right) A^2+\left(\frac{12608}{1125}-\frac{4 \pi ^2}{5}\right) A-\frac{\pi ^3}{16}+\frac{3 \pi }{8}\right] \frac{\dot{b}^2}{b^4}
    + \left(\frac{64}{75}-\frac{5 \pi  A}{16}\right) \frac{\ddot{A}}{b^2}\\
    + \left( \frac{5 \pi  A^2}{16}+\frac{16 A}{15}-\frac{\pi}{4} \right) \frac{1}{b^2}
    + \left[\left(\frac{\pi ^3}{3}-\frac{275 \pi }{144}\right) A^2+\left(\frac{8 \pi ^2}{15}-\frac{25216}{3375}\right) A+\frac{\pi ^3}{24}-\frac{\pi}{4}\right] \frac{\ddot{b}}{b^3} \\
    +2 \pi  A^2
    +\left[\left(\frac{2 \pi ^3}{3}-\frac{275 \pi }{72}\right) A+\frac{8 \pi ^2}{15}-\frac{25216}{3375}\right] \frac{\dot{A} \dot{b}}{b^3}+\frac{16 A}{15}+\frac{\pi }{4} = 0
\end{multline*}

\begin{multline*}
    \frac{5 \pi  \ddot{A}}{8 b}
    + \left[\left(\frac{365 \pi }{144}-\frac{\pi ^3}{3}\right) A+\frac{6848}{3375}-\frac{4 \pi^2}{15}\right] \frac{\dot{b}^2}{b^3}\\
    +\left(\frac{64}{75}-\frac{5 \pi  A}{16}\right) \frac{\ddot{b}}{b^2}
    -\frac{5 \pi  \dot{A} \dot{b}}{8 b^2}+A \left(4 \pi  b-\frac{5 \pi }{8 b}\right)+\frac{16 b}{15}-\frac{16}{15 b} = 0
\end{multline*}

\section{Perturbation expressions}
\label{app:partial-solutions}

\begin{align*}
&\chi_{R_1}(t,x)=(x-t)^2+\left(\frac{\sqrt{7}\epsilon}{4} -\sqrt{2}\right) (x-t) +\frac{1}{192}
\left(96-24 \sqrt{14} \epsilon +25 \epsilon ^2\right)\\
&\chi_{R_2}(t,x)=-\frac{5}{2}(x-t)^2+\left(\frac{5}{\sqrt{2}}-\frac{\sqrt{7} \epsilon }{8}\right)(x-t)+\frac{1}{384}
\left(-480+24 \sqrt{14} \epsilon +23 \epsilon ^2\right)\\
&\chi_{R_3}(t,x)=(x-t)^2-\left(\sqrt{2}+\frac{\sqrt{7} \epsilon }{4}\right)(x-t)+\frac{1}{192} \left(96+24
\sqrt{14} \epsilon +13 \epsilon ^2\right)\\
&\chi_{R_4}(t,x)=\frac{5}{2}(x-t)^2-\left(\frac{5}{\sqrt{2}}+\frac{3 \sqrt{7} \epsilon }{8}\right)(x-t)+\frac{1}{128}
\left(160+24 \sqrt{14} \epsilon +11 \epsilon ^2\right)\\
&\chi_{R_5}(t,x)=-(x-t)^2+\left(\sqrt{2}+\frac{\sqrt{7} \epsilon }{4}\right)(x-t)+\frac{1}{64} \left(-32-8
\sqrt{14} \epsilon -7 \epsilon ^2\right)
\end{align*}

\begin{align*}
&\chi_{L_1}(t,x)=-\frac{1}{7}(x+t)^2+\frac{1}{28} \left(4 \sqrt{2}+\sqrt{7} \epsilon \right)(x+t)+\frac{-96-24
\sqrt{14} \epsilon +7 \epsilon ^2}{1344}\\
&\chi_{L_2}(t,x)=\eta_{L_1}(t,x)\\
&\chi_{L_3}(t,x)=\frac{1}{14}(x+t)^2+\frac{1}{56} \left(-4 \sqrt{2}+\sqrt{7} \epsilon \right)(x+t)+\frac{1}{896}
\left(32-8 \sqrt{14} \epsilon +7 \epsilon ^2\right)\\
&\chi_{L_4}(t,x)=\eta_{L_3}(t,x)\\
&\chi_{L_5}(t,x)=\eta_{L_3}(t,x)
\end{align*}

\bibliography{bibliography}

\end{document}